\documentclass[11pt,english]{article}
\usepackage{jheppub}

\usepackage[T1]{fontenc}
\usepackage[latin9]{inputenc}
\usepackage{amsmath}
\usepackage{esint}
\usepackage{babel}
\usepackage{color}
\usepackage{enumitem}
\usepackage{subfigure}

\usepackage[numbers,sort&compress]{natbib}
\usepackage{hyperref}
\allowdisplaybreaks
\hypersetup{
colorlinks=true,
linkcolor=blue,
linktocpage=true,
citecolor=violet
}

\usepackage{xargs}
\usepackage[colorinlistoftodos,prependcaption,textsize=tiny]{todonotes}

\def\dlog{{$d\log$}}
\def\eps{\epsilon}

\def\Mathematica{{{\sc Mathematica}}}
\def\FeynArts{{{\sc FeynArts}}}
\def\FeynCalc{{{\sc FeynCalc}}}
\def\FeynRules{{{\sc FeynRules}}}
\def\Dlog{{{\sc Dlog}}}
\def\MultivariateApart{{{\sc MultivariateApart}}}
\def\Reduze{{{\sc Reduze}}}

\def\Sam{{{\sc S@M}}}
\def\Tam{{{\sc T@M}}}

\newcommand{\munich}{Max-Planck-Institut f\"ur Physik, Werner-Heisenberg-Institut, 
80805 M\"unchen, Germany.}

\begin{document}
\preprint{MPP-2021-209}

\title{Maximal transcendental weight contribution of scattering amplitudes}

\author[a]{Johannes M. Henn}
\author[a]{and William J. Torres~Bobadilla}
\affiliation[a]{\munich}
\emailAdd{henn@mpp.mpg.de}
\emailAdd{torres@mpp.mpg.de}

\abstract{
Feynman integrals in quantum field theory evaluate to special functions and numbers that are usefully described by the notion of transcendental weight. In this paper, we propose a way of projecting a given dimensionally-regularised Feynman integral, for example contributing to a scattering amplitudes, onto its maximal weight part. The method uses insights into the singularity structure of space-time loop integrands, and is complementary to usual generalised unitarity approaches. We describe the method and give a proof-of-principle application to the two-loop scattering amplitudes $gg \to H$ in the heavy top-quark mass limit, which involves both planar and non-planar Feynman integrals. We also comment on further possible applications and discuss subtleties related to evanescent integrand terms.
}

\maketitle

\graphicspath{{figs/}}

\section{Introduction}
\label{sec:intro}

Observables computed in quantum field theory often depend on interesting transcendental numbers and special functions that arise from Feynman integrals. Understanding the structure of the perturbative expansion, and finding efficient ways of computing it are important challenges. 

In practice, the integrated answers can be organised according to the complexity of special functions and numbers appearing in them. For example, in anomalous dimensions one encounters, among others, (multiple) zeta values $\zeta_n = \sum_{k\ge 1} 1/k^n$. 
This concept also extends naturally to functions written in terms of iterated integrals, 
see e.g.~\cite{Goncharov:2010jf}.
At a given loop order, there is a heuristic bound for the maximal value of $n$, also referred to as transcendental weight, or `transcendentality'~\cite{Kotikov:2004er}, 
namely $2L$ for $L$-loop integrals in four dimensions 
(cf.~\cite{Hannesdottir:2021kpd} and references therein).
A remarkable conjecture states that for certain quantities, the maximal weight terms agree between the maximally supersymmetric Yang-Mills theory ($\mathcal{N}=4$ sYM) and quantum chromodynamics (QCD)~\cite{Kotikov:2004er}. Maximal weight refers to the terms with $n=2 L$, 
where $L$ is the loop order.
This observation was instrumental in obtaining three-loop anomalous dimensions of twist-two anomalous dimensions in $\mathcal{N}=4$ sYM~\cite{Kotikov:2004er} from the QCD results. More recently, it was used to predict the planar four-loop value of the planar cusp anomalous dimension in $\mathcal{N}=4$ sYM~\cite{Dixon:2017nat}.

Unfortunately, the one-to-one correspondence between maximal weight terms in $\mathcal{N}=4$ sYM and QCD is restricted to certain observables only. In the original case, it can be argued that it comes about thanks to special properties of the DGLAP evolution equation that governs twist-two anomalous dimensions~\cite{Kotikov:2002ab}, 
namely that the gluonic contributions give the relevant contributions in this case. 
For more general observables, this does not seem to be the case, 
and, therefore, the maximal weight terms can differ between $\mathcal{N}=4$ sYM and QCD. 
Nevertheless, one can ask: is there a good way of obtaining the maximal weight contributions of an observable? 

Insight into this question comes from studies in $\mathcal{N}=4$ sYM. It was noticed, heuristically, that observables computed in this theory are expressed in terms of maximal weight functions (see e.g. the review~\cite{Henn:2020omi} and references therein). It turns out that this property can be anticipated by looking at the types of Feynman integrals that appear in that theory~\cite{Arkani-Hamed:2010pyv,Arkani-Hamed:2014via}. 
What was noticed is that integrals in this theory seem to have the special property that their integrands can be written as a `\dlog{}' form. 
In particular, this means that they do not have any double pole on any (generalised unitarity) cuts. 
This observation was crucial, among other things, to bootstrap integrands of the theory, in combination with generalised unitarity or related techniques. In some cases, the \dlog{} property can be proven via loop level recursion relations~\cite{Arkani-Hamed:2010zjl}. 

Although initially observed in $\mathcal{N}=4$ sYM, 
the idea of \dlog{} integrands and their connection to maximal weight functions turned out to be dramatically useful in the computation of generic Feynman integrals. 
In the state-of-the-art approach, one uses integration-by-parts (IBP) 
relations~\cite{Tkachov:1981wb,Chetyrkin:1981qh,Laporta:2001dd} 
to write scattering amplitudes in terms of a basis of Feynman integrals, called master integrals (MIs), 
where a major problem and bottleneck is in their computation.
MIs are known to satisfy (a system of) differential equations, but the latter is typically complicated, and depends on the set of kinematic invariants and on the dimensional regulator $\eps$, where e.g. $D=4-2\eps$. Moreover, the form of the differential equations depends on the choice of integral basis.
A key observation is that if a basis of maximal (and uniform) weight integrals is chosen, then the differential equations take a simple, canonical form~\cite{Henn:2013pwa}. In that form, the differential equations can be solved easily in a series expansion in the dimensional regulator $\eps$. These ideas have led to numerous new calculations relevant for phenomenology.  

In this paper, we build on the insights of loop integrands. We wish to understand directly at integrand level what terms lead to the `maximal weight' part of the answer, avoiding the (often computationally heavy) IBP machinery. To do this, we propose a way of systematically extracting the `maximal weight' contributions of Feynman integrands, without the need to evaluate the integrals. 

The procedure we propose is the following. For a given scattering process, one first considers the denominator structure of the Feynman integrals. Firstly, for each denominator structure, one determines the most general set of \dlog{} integrands that can be written down. This is an algorithmic procedure, in principle. Secondly, one imagines a decomposition of the original integrand in terms of the \dlog{} basis just constructed, plus other terms. The latter have, by construction, at least a double pole, and are therefore expected to produce less than maximal weight terms only. The third step then consists in projecting the amplitude onto the \dlog{} basis only. We show how the coefficients of these integrals are obtained via a residue calculation (making use of algebraic geometry techniques).
In this way, we define a maximal weight projection of the original integrand. 
The method we propose bears 
several connections and was in part inspired by unitarity-based methods~\cite{Gluza:2010ws,Caron-Huot:2012awx,Ossola:2006us,Mastrolia:2011pr,Zhang:2012ce,Badger:2013gxa,Mastrolia:2012an,Mastrolia:2012wf,Sogaard:2013fpa,Mastrolia:2016dhn,Mastrolia:2016czu,Ita:2015tya,Bourjaily:2017wjl,Bourjaily:2019iqr,Bourjaily:2020qca,Bourjaily:2021hcp}, and can be considered as a complementary tool. 

In what situations do we expect the procedure we propose to apply?
Let us discuss two possible caveats. The first one is related to the class of functions that can appear in Feynman integrals, and the second one has to do with dimensional regularisation~\cite{tHooft:1972tcz,Bollini:1972ui}.
Let us discuss these two issues in turn. 

Firstly, we expect the method to work as described in cases where the integrals evaluate to multiple poly-logarithms, for which the concept of transcendental weight is well understood (at least, to the satisfaction of physicists).
For integrals evaluating to more general transcendental functions, such as elliptic poly-logarithms and beyond, modifications are likely to be needed, which however go beyond the scope of this work.

Secondly, in practice it is often desirable to compute Feynman integrals in dimensional regularisation, with e.g. $D=4-2\eps$.
Of course, it is much easier to perform the integrand analysis in integer dimensions, and often this is in fact sufficient (see e.g. a discussion in~\cite{Henn:2014qga}). The distinction between integer and non-integer   dimension can become relevant in cases that involve many independent momenta, because then certain Gram determinants vanish in e.g. $D=4$ dimensions, but not in $D=4-2\eps$. See e.g.~\cite{Chicherin:2018old}, where a more refined~\dlog{} analysis is performed, using Baikov representations~\cite{Baikov:1996rk,Frellesvig:2017aai,Zeng:2017ipr,Bosma:2017hrk,Dlapa:2021qsl}.
For the purposes of the present paper, we wish to restrict ourselves to a four-dimensional integrand analysis.
This means that we tacitly assume that maximal weight terms can be extracted from the knowledge of naive four-dimensional integrands.
Since this may not be the case in all situations, we therefore study this important point in this paper. We show that if one is interested in knowing one-loop amplitudes up-to and including the finite part only, then considering naive four-dimensional integrands is sufficient. 
On the other hand, the same is not true for evanescent terms in the dimensional regulator --- those terms may contribute maximal weight terms at five points, for example. The latter affect the analysis at higher loops, so that one can hope to extract the maximal weight terms there only when considering appropriately renormalised observables. We discuss explicit examples of these features for two-loop five-point scattering amplitudes.

In order to validate our method, we illustrate it for a two-loop scattering process in QCD, namely $H\to gg$ amplitudes at the two-loop order. In order to do so, we first generate all Feynman diagrams, which include non-planar cases. We then construct an explicit basis of \dlog{} integrands for all relevant Feynman diagram topologies. With this basis at our disposal, we project the general integrands onto our \dlog{} basis. This new expression is expected to give the maximal weight part of the QCD process. We test this hypothesis by explicitly comparing the original and the projected expression after an IBP reduction. We also plug in the explicit values of MIs to illustrate the effect of the projection at the level of the $\eps$-expansion. \\

The outline of this paper is as follows.
Section~\ref{sec:algebraic} contains a review of \dlog{} representations, and discusses the representations that will be needed for the purpose of this paper. 
In section~\ref{sec:decodlog}, we describe in more detail the idea of the maximal weight projection, before applying it in section~\ref{sec:applications} to $H\to gg$ amplitudes at the two-loop order. 
Then, in section~\ref{sec:evanescent}, we discuss in which cases subtleties with evanescent terms can be expected, and discuss ideas on avoiding them. 
Finally, we give our conclusion and outlook in section~\ref{sec:conclusions}.

\section{\dlog{} basis for $H\to gg$ integrands at two loops}
\label{sec:algebraic}

In this section, we briefly review how to find Feynman integrands that admit a \dlog{} representation in four dimensions. 
Many examples of such Feynman integrals can be found in the literature, especially at one loop, 
see e.g.~\cite{Bern:2014kca}. This provides very useful information for constructing the higher-loop \dlog{} representations that we will need for the $H\to gg$ that we wish to study. 
The reason is that in many cases, one can obtain a higher-loop representation using a loop-by-loop approach. The representations obtained in this way have the additional advantage that they tend to involve few terms, and often can be written in terms of single terms only.

In order to consistently apply the maximal weight projection method, it is important that we classify {\it all} \dlog{} integrands for a given Feynman integral topology, as otherwise we might miss relevant contributions. In order to address this issue, we can profit from an algorithmic method for identifying \dlog{} integrands \cite{Wasser:2018qvj,Henn:2020lye}.
Ref.~\cite{Henn:2020lye} also provides a public implementation, the \Mathematica{} package \Dlog{}.
This algorithm takes as input a given Feynman diagram topology (i.e., the propagator structure) and considers, as starting point, 
an ansatz for the most general numerator that can lead to \dlog{} integrands.
The ansatz is finite because the numerator is restricted to obey certain power counting rules~\cite{Henn:2020lye}.
The algorithm then employs a convenient parametrisation of the four-dimensional loop momenta, similar to what we will use later in this section.
The resulting integrand is a rational function in $4L$ integration variables, and depends on the external kinematics, and on the parameters of the ansatz. The algorithm then analyses, iteratively, residues that can be taken in each integration variable, until one arrives at the leading singularities, i.e. the maximal residues of the integrand. Whenever a double or higher pole is encountered, a condition is placed on the ansatz parameter to nullify the expression.
In this way, a complete list of \dlog{} integrands for a given integral topology is obtained.

In this way, the algorithm~\cite{Henn:2020lye} informs us of the total number of \dlog{} integrals that we need to consider, and it also suggests a basis for them. In what follows, we also need to know the explicit \dlog{} representation. This feature has not been implemented in the \Dlog{} procedure for performance reasons, but could be incorporated easily. 
However, a typical \dlog{} representation would involve many terms, even for simple integrands, whereas we know that often, one can find a one-term \dlog{} representation, which is clearly advantageous. For this reason, we prefer, for the present purpose, 
to construct more compact \dlog{} forms by the loop-by-loop approach alluded to earlier.

Thus, we begin this section with the study of 
one-loop \dlog{} integrands that we will need later in our two-loop analysis.
It turns out that the classes of integrands we need contain
either Feynman or eikonal propagators. 

\subsection{One-loop \dlog{} integrands}
\label{sec:1Ldlog}

Let us illustrate the procedure for constructing \dlog{} representations for one-loop Feynman integrands. 
We do so for convenience of readers, and in order to keep this paper self-contained. 
Of course, similar discussions can be found in the literature, see e.g.~\cite{Bern:2014kca,Wasser:2018qvj}.

As was mentioned earlier, we take the loop momentum to be four-dimensional for this analysis.
It is useful to choose a convenient parametrisation in a 
four-dimensional basis $\mathcal{E}=\{e_1,e_2,e_3,e_4\}$, 
such that we can write
\begin{align}
k_{i}^{\mu}=\alpha_{i,1}\,e_{1}^{\mu}+\alpha_{i,2}\,e_{2}^{\mu}+\alpha_{i,3}\,e_{3}^{\mu}+\alpha_{i,4}\,e_{4}^{\mu}\,.
\label{eq:LoopParam}
\end{align}
If there are at least two independent massless momenta, say $p_1$ and $p_2$.
In this case, a particularly convenient choice of the basis vectors is as follows,
\begin{align}\label{eq:LoopParam2}
e_1^\mu=p_1^\mu\,,\quad e_2^\mu=p_2^\mu \,,\quad
e_3^\mu=\langle1|\gamma^\mu|2]/2\,, \quad
e_4^\mu=\langle2|\gamma^\mu|1]/2 \,.
\end{align}
Here, angle and square brackets refer to the spinor-helicity formalism~\cite{Dixon:1996wi}.

\begin{figure}[t]
\centering
\subfigure[]{\label{fig:int_fam_4pt1L-a}\includegraphics[scale=0.67]{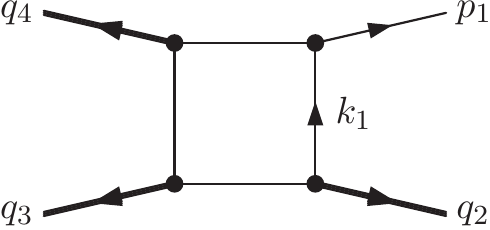}}\:
\subfigure[]{\label{fig:int_fam_4pt1L-b}\includegraphics[scale=0.67]{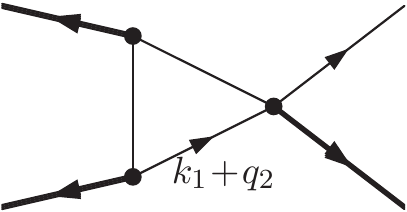}}\:
\subfigure[]{\label{fig:int_fam_4pt1L-c}\includegraphics[scale=0.67]{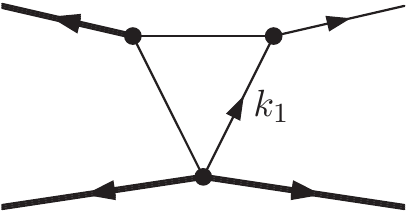}}\:
\subfigure[]{\label{fig:int_fam_4pt1L-d}\includegraphics[scale=0.67]{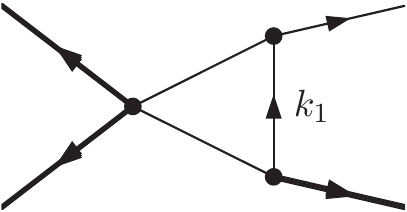}}\:
\subfigure[]{\label{fig:int_fam_4pt1L-e}\includegraphics[scale=0.67]{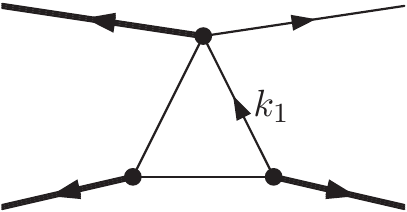}}
\caption{\dlog{} basis for integrand family~\eqref{eq:1Lprops}.}
\label{fig:int_fam_4pt1L}
\end{figure}
Let us apply this method to the four-point integrals shown in Fig.~\ref{fig:int_fam_4pt1L-a}.
It stands for an integrand
the ones constructed only from Feynman propagators and 
consider the family of integrands generated from the following set of 
propagators,
\begin{align}
{\cal I} = \frac{d^4k_1\, {\cal N}}{  \left(k_{1}-p_{1}\right)^{2} k_{1}^{2} \left(k_{1}+q_{2}\right)^{2} \left(k_{1}+q_{2}+q_{3}\right)^{2} } \,.
\label{eq:1Lprops}
\end{align}
where $p_i^2=0$ and $q_{i}$ can be
considered as massless ($q_i=p_i$) or massive ($q_{i}^2\ne 0$), depending on the case we wish to consider,
with the momentum conservation $p_1+q_2+q_3+q_4=0$. 
Therefore, the kinematic invariants can be cast as, $s=(p_1+q_2)^2$ and $t=(q_2+q_3)^2$. 

\subsubsection*{\dlog{} forms for one-loop integrands with massless propagators}

Given the integrand~\eqref{eq:1Lprops}, 
and the parametrisation~\eqref{eq:LoopParam}, we would now like to
generate a \dlog{} basis for the family of integrands~\eqref{eq:1Lprops}.
To this end, we rely on the \Mathematica{} package \Dlog{},
where, after using its built-in routines, 
we find a set of five one-loop \dlog{} integrands.
This set of \dlog{} integrands, depicted in Fig.~\ref{fig:int_fam_4pt1L},
consists of one scalar box, and four scalar triangles. 

Having found the \dlog{} basis, let us now illustrate how one can write down an explicit \dlog{} representation.
For example, consider the triangle integral depicted in Fig.~\ref{fig:int_fam_4pt1L-d}, 
where for simplicity we take $q_2 = p_2$ to be massless,
\begin{align}
\mathcal{I}=\frac{d^{4}k_{1}}{\left(k_{1}-p_{1}\right)^{2}k_{1}^{2}\left(k_{1}+p_{2}\right)^{2}}\,.
\label{eq:sca_tri_alpha}
\end{align}
A short calculation shows that, upon plugging in the
 parametrisations~\eqref{eq:LoopParam},~\eqref{eq:LoopParam2}, 
and taking the Jacobian factor into account, one obtains,
\begin{align}
\mathcal{I} = \frac{d\alpha_{1}\,d\alpha_{2}\,d\alpha_{3}\,d\alpha_{4}}{s\left(\left(\alpha_{1}-1\right)\alpha_{2}-\alpha_{3}\alpha_{4}\right)\left(\alpha_{1}\alpha_{2}-\alpha_{3}\alpha_{4}\right)\left(\alpha_{1}\left(\alpha_{2}+1\right)-\alpha_{3}\alpha_{4}\right)}
\,.
\end{align}
where we have dropped an irrelevant (because kinematic-independent) overall factor.
Notice that, to simplify the notation in this section, we define 
the loop components of the loop momentum parametrisation~\eqref{eq:LoopParam} 
as $\alpha_{1,i} = \alpha_{i}$. 

The integrand~\eqref{eq:sca_tri_alpha} contains three denominator factors, and it appears natural to change variables $\alpha_{1,2,3} \to \tau_{1,2,3}$ according to  
\begin{align}
\tau_{1}=&\left(\left(\alpha_{1}-1\right)\alpha_{2}-\alpha_{3}\alpha_{4}\right)\,s\,,
\notag\\
\tau_{2}=&\left(\alpha_{1}\alpha_{2}-\alpha_{3}\alpha_{4}\right)\,s\,,
\notag\\
\tau_{3}=&\left(\alpha_{1}\left(\alpha_{2}+1\right)-\alpha_{3}\alpha_{4}\right)\,s\,.
\end{align}
Remarkably, upon doing so, one finds the following simple expression,
\begin{align}
\mathcal{I}&=\frac{1}{s}d\log\tau_{1}\,d\log\tau_{2}\,d\log\tau_{3}\,d\log\alpha_{4}\,.
\label{triangletau}
\end{align}
At this stage we emphasise that it is remarkable that Eq.~\eqref{triangletau} consists of a single term only.
We will in fact find similar forms for the other integrals considered in this paper as well.
To appreciate that this is by no means guaranteed, let us write an equivalent expression in the $\alpha_i$ variables,
\begin{align}
\omega^{\text{1m}-\text{tri}}\left(k_{1};p_{1},p_{2}\right)  =&
-d\log\left(\alpha_{4}\right)d\log\left(\alpha_{2}\right)d\log\left(\alpha_{3}\right)d\log\left(\alpha_{1}\alpha_{2}-\alpha_{3}\alpha_{4}\right)
\notag\\
&-d\log\left(\alpha_{4}\right)d\log\left(\alpha_{2}\right)d\log\left(\alpha_{2}^{2}+\alpha_{2}+\alpha_{3}\alpha_{4}\right)d\log\left[\alpha_{1}\left(\alpha_{2}+1\right)-\alpha_{3}\alpha_{4}\right]
\notag\\
&+d\log\left(\alpha_{4}\right)d\log\left(\alpha_{2}\right)d\log\left(\alpha_{2}^{2}+\alpha_{2}+\alpha_{3}\alpha_{4}\right)d\log\left(\alpha_{1}\alpha_{2}-\alpha_{2}-\alpha_{3}\alpha_{4}\right)
\notag\\
&+d\log\left(\alpha_{4}\right)d\log\left(\alpha_{2}\right)d\log\left(\alpha_{3}\right)d\log\left[\alpha_{1}\left(\alpha_{2}+1\right)-\alpha_{3}\alpha_{4}\right]\,.
\label{triangledlogalternative}
\end{align}
 Having made this comment, let us return to the simpler representation of Eq.~\eqref{triangletau}.
Taking into account the explicit definitions for $\tau_i$ and $\alpha_4$
in terms of denominators and scalar products, 
we note that this integrand can be re-expressed as follows, 
\begin{align}
\mathcal{I}&=\frac{1}{s}d\log\left(k_{1}-p_{1}\right)^{2}\,d\log k_{1}^{2}\,d\log\left(k_{1}+p_{2}\right)^{2}\,d\log\left(2\,k_{1}\cdot e_{3}\right)\,.
\end{align}
In this form it is clear that the overall factor $1/s$ corresponds 
to the four-dimensional leading singularity
of the scalar triangle $\mathcal{I}$. 
Hence, accounting for the latter, we can define the \dlog{} integrand, 
\begin{align}
\mathcal{I}\left(\parbox{24mm}{\includegraphics[scale=0.5]{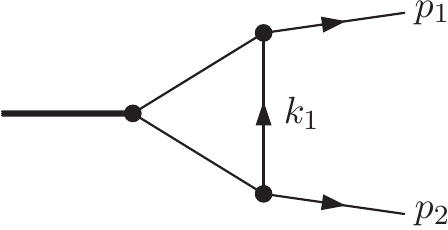}}\,s\right)
&\equiv 
\omega^{\text{1m}-\text{tri}}\left(k_1;p_{1},p_{2}\right)\,,
\label{eq:1mass_tri_pic}
\end{align}
with, 
\begin{align}
\omega^{\text{1m}-\text{tri}}\left(k_1;p_{1},p_{2}\right)
& = d\log\left(k_{1}-p_{1}\right)^{2}\,d\log k_{1}^{2}\,d\log\left(k_{1}+p_{2}\right)^{2}\,d\log\left(2\,k_{1}\cdot e_{3}\right)\,.
\label{eq:1mass_tri_expr}
\end{align}
In Eq.~\eqref{eq:1mass_tri_pic}, 
the l.h.s.,  $\mathcal{I}(\cdots)$ stands for the integrand of the loop topology, 
whose Feynman propagators can be read off from the diagram.
Thus, we have found a \dlog{} form $\omega^{\text{1m}-\text{tri}}\left(k_1;p_{1},p_{2}\right)$
for the scalar ``one-mass triangle'' integral.
In the pictorial representation of integrands, 
here and in the following, 
external thin and thick lines represent, respectively, massless (on-shell) and massive (or off-shell) momenta.

As illustrated for the scalar triangle in Eq.~\eqref{eq:sca_tri_alpha},
finding an explicit \dlog{}
representation for a given integrand family, say~\eqref{eq:1Lprops}, 
is a procedure at integrand level. 
Relations that arise as consequence of dimensional regularisation
at integral level are not taken into account at this stage.
For instance, in the case where all external momenta are massless, we 
have five \dlog{} basis integrands, but upon using IBP relations, only three
of them are independent.
In fact, it is important for the consistency of our method that a complete 
integrand basis is constructed.

Let us return to the construction of one-loop \dlog{} representations.
Following the same strategy adopted for the one-loop scalar triangle, 
we find algebraic expressions for the following \dlog{} forms 
(which correspond to different kinematic configurations of integrals Fig.~\ref{fig:int_fam_4pt1L}).
For the two-mass triangle, we have
\begin{align}\label{eq:twomasstriangle}
\omega^{\text{2m}-\text{tri}}\left(k_1;p_{1},q_{2}\right) \equiv&
\mathcal{I}\left(\parbox{24mm}{\includegraphics[scale=0.5]{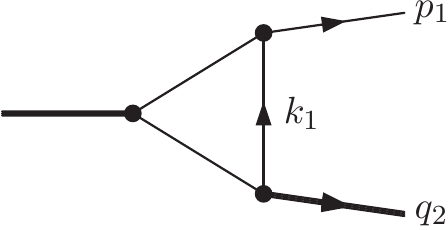}}
\left(2\,p_1\cdot q_2\right)
\right) \\
 =& d\log\left(k_1-p_{1}\right)^{2}\,d\log k_1^{2}\,d\log\left(k_1+q_{2}\right)^{2}\,d\log\left(2k_1\cdot e_{3}\right)\,,\nonumber
\end{align}
For the ``two-mass-hard'' and ``two-mass-easy'' boxes, we have~\cite{Bern:2014kca},
\begin{align}\label{eq:twomasshardbox}
\omega^{\text{2mh}-\text{box}}\left(k_1;p_{1},p_{2},q_{3}\right) &\equiv
\mathcal{I}\left(\parbox{24mm}{\includegraphics[scale=0.5]{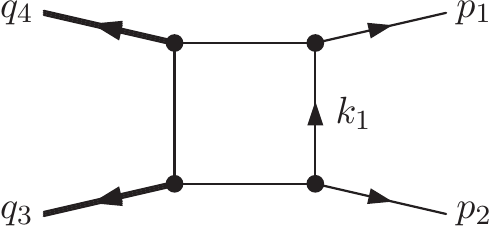}}\,s\,t\right) \\
&\hspace{-2cm}= \mp d\log\frac{\left(k_1-p_{1}\right)^{2}}{\left(k_1-k^{\pm}_1\right)^{2}}\,d\log\frac{k^{2}_1}{\left(k_1-k^{\pm}_1\right)^{2}}\,d\log\frac{\left(k_1+p_{2}\right)^{2}}{\left(k_1-k^{\pm}_1\right)^{2}}\,d\log\frac{\left(k_1+p_{2}+q_3\right)^{2}}{\left(k_1-k^{\pm}_1\right)^{2}}\,,\nonumber
\end{align}
and
\begin{align}\label{eq:twomasseasybox}
\omega^{\text{2me}-\text{box}}\left(k_1;p_{1},q_{2},p_{3}\right) 
&\equiv
\mathcal{I}\left(\parbox{24mm}{\includegraphics[scale=0.5]{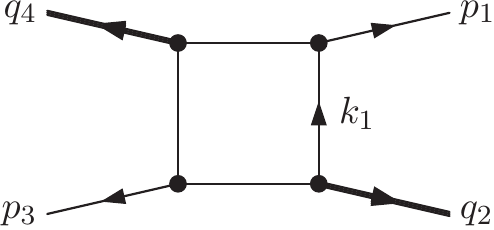}}\,
\left(s\,t-q_{2}^{2}\,q_{4}^{2}\right)
\right)
\,, \\
&\hspace{-2cm}=
\mp d\log\frac{\left(k_1-p_{1}\right)^{2}}{\left(k_1-k^{\pm}_1\right)^{2}}d\log\frac{k^{2}_1}{\left(k_1-k^{\pm}_1\right)^{2}}d\log\frac{\left(k_1+q_{2}\right)^{2}}{\left(k_1-k^{\pm}_1\right)^{2}}d\log\frac{\left(k_1+q_{2}+p_3\right)^{2}}{\left(k_1-k^{\pm}_1\right)^{2}}\,,\nonumber
\end{align}
respectively.
Here $k^{\pm}$ are the two solutions of the maximal cut conditions,
\begin{align}
\left(k_1^{\pm}-p_{1}\right)^{2}=\left(k^{\pm}_1\right)^2
=\left(k_1^{\pm}+q_{2}\right)^{2}=\left(k_1^{\pm}+q_{2}+q_{3}\right)^{2}=0\,.
\end{align}

Let us emphasise that the additional prefactors that appears in the definition
of integrands correspond to four-dimensional leading singularities 
of the scalar Feynman integrand. 
Also, let us note that although the \dlog{} form 
of the four-point Feynman integrands involves the auxiliary factor 
$\left(k_1-k^{\pm}_1\right)^{2}$, the integrand does not have a residue at $k_1=k_1^{\pm}$.
We refer interested readers to~\cite{Bern:2014kca} for more details.

\subsubsection*{\dlog{} forms for integrands with eikonal and Feynman propagators}

We shall see in Sec.~\ref{sec:2Ldlog} that when following a loop-by-loop approach for writing down \dlog{} integrands, 
additional  one-loop \dlog{} forms are needed.
The required integrands containing both standard massless Feynman propagators, and eikonal propagators. 
The latter are factors of the form $1/(2k\cdot p_i)$.
Hence, to complete the set of building blocks needed to generate the two-loop
 \dlog{} forms in Sec.~\ref{sec:2Ldlog}, we presently provide their explicit algebraic expressions.

We have,
\begin{align}\label{eq:triangleeikonal1dlog}
\omega_{\text{Eikonal 1}}^{\text{1m}-\text{tri}}\left(k_1;p_{1},p_{2}\right) &\equiv
\mathcal{I}\left(\parbox{24mm}{\includegraphics[scale=0.5]{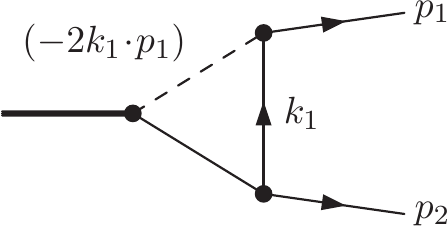}}\,s\right) \\
&=d\log\left(-2k_1\cdot p_{1}\right)\,d\log k_1^{2}\,d\log\left(k_1+p_{2}\right)^{2}\,d\log\left(2k_1\cdot e_{3}\right)\,, \nonumber
\end{align}
\begin{align}\label{eq:triangleeikonal2dlog}
\omega_{\text{Eikonal 2}}^{\text{1m}-\text{tri}}\left(k_1;p_{1},p_{2}\right)&\equiv
\mathcal{I}\left(\parbox{24mm}{\includegraphics[scale=0.5]{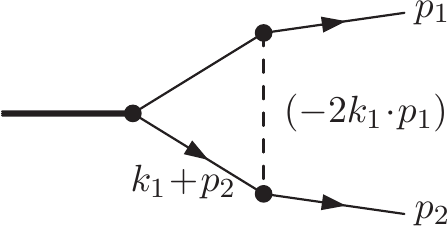}}\,s\right)\\
&= d\log\left(-2k_1\cdot p_{1}\right)\,d\log\left(k_1-p_{1}\right)^{2}\,d\log\left(k_1+p_{2}\right)^{2}\,d\log\left(2k_1\cdot e_{3}\right)\,,\nonumber
\end{align}
\begin{align}\label{eq:boxeikonal1dlog}
\omega_{\text{Eikonal 1}}^{\text{Box}}\left(k_1;p_{1},p_{2}\right)&\equiv
\mathcal{I}\left(\parbox{24mm}{\includegraphics[scale=0.5]{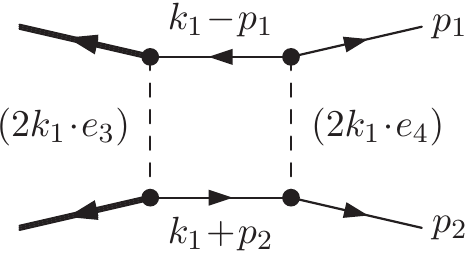}}\,s^2\right)\\
&=d\log\frac{2k_1\cdot e_{3}}{k^{2}_1}\,d\log\frac{2k_1\cdot e_{4}}{k^{2}_1}\,d\log\frac{\left(k_1-p_{1}\right)^{2}}{k^{2}_1}\,d\log\frac{\left(k_1+p_{2}\right)^{2}}{k^{2}_1}\,,\nonumber
\end{align}
\begin{align}\label{eq:boxeikonal2dlog}
\omega_{\text{Eikonal 2}}^{\text{Box}}\left(k_1;p_{1},p_{2}\right) &\equiv
\mathcal{I}\left(\parbox{24mm}{\includegraphics[scale=0.5]{eik_box1L_num.pdf}}\,s\,k^2_1\right)\\
&=d\log\frac{\left(k_1-p_{1}\right)^{2}}{2k_1\cdot e_{4}}d\log\frac{\left(k_1+p_{2}\right)^{2}}{2k_1\cdot e_{3}}d\log\frac{2k_1\cdot p_{1}}{2k_1\cdot e_{4}}d\log\left(2k_1\cdot e_{3}\right)\,. \nonumber
\end{align}
Notice that in the pictorial representation of the integrands, eikonal propagators
are depicted as dashed lines.
To make the definition unambiguous, their explicit denominator factor, e.g. $(2p_1\cdot k)$, is 
printed next to the line. 
For instance, we have explicitly,
\begin{align}
\mathcal{I}\left(\parbox{24mm}{\includegraphics[scale=0.5]{eik_tri1L_1.pdf}}\,s\right) =  s \frac{d^4 k_1}{(-2k_1\cdot p_{1}) k^{2}_1 \left(k_1+p_{2}\right)^{2}}  \,.
\end{align}

\subsection{Two-loop \dlog{} integrands}
\label{sec:2Ldlog}

Let us now turn our attention to the calculation 
of algebraic expressions for two-loop \dlog{} integrands. 
A procedure similar to the one presented in the previous 
section could be carried out, namely to parametrise  the loop 
momenta, $k_1$ and $k_2$, according to~\eqref{eq:LoopParam}
and to look for the explicit algebraic expressions in terms of 
products of \dlog{} forms. 
However, to profit from the simple algebraic expressions obtained 
for one-loop \dlog{} integrands,
one can recycle and use them in the calculation of \dlog{}
integrands at higher loops. 
In this section, we follow such a loop-by-loop approach,
where one-loop \dlog{} integrands are the needed building blocks 
for the generation of multi-loop ones. 
We explicitly provide the derivation of 
planar and non-planar \dlog{} three-point Feynman integrands at two loops
and also discuss representative four-point \dlog{} forms.


\begin{figure}
\centering
\subfigure[]{\label{fig:parent_topos_Hgg-a}\includegraphics[scale=0.67]{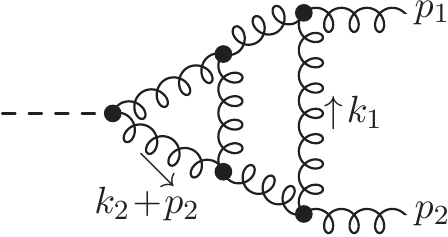}}\:
\subfigure[]{\label{fig:parent_topos_Hgg-b}\includegraphics[scale=0.67]{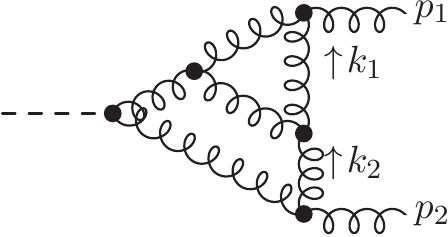}}\:
\subfigure[]{\label{fig:parent_topos_Hgg-c}\includegraphics[scale=0.67]{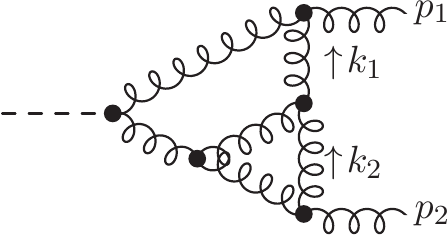}}\:
\subfigure[]{\label{fig:parent_topos_Hgg-d}\includegraphics[scale=0.67]{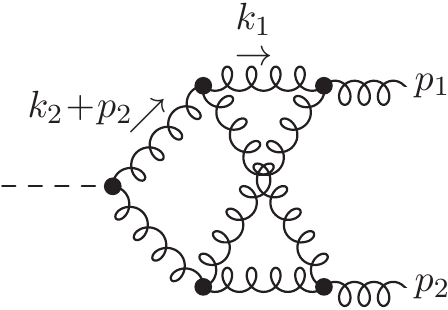}}
\caption{Representative two-loop (parent) topologies with seven propagators present 
in the scattering amplitude $H\to gg$.}
\label{fig:parent_topos_Hgg}
\end{figure}

We are interested in finding a \dlog{} basis for the process $H\to gg$ at two loops.
This involves the generation of planar and non-planar Feynman diagrams, 
as shall be illustrated in detail in section~\ref{sec:applications}.
The complete set of Feynman diagrams contributing to this scattering amplitude
can be grouped into four `parent' topologies with six propagators, as depicted in 
Fig.~\ref{fig:parent_topos_Hgg}. 
Let us discuss them in turn.

\begin{figure}[t]
\begin{center}
\subfigure[]{\label{fig:2L_dlog_tri-a}\includegraphics[scale=0.67]{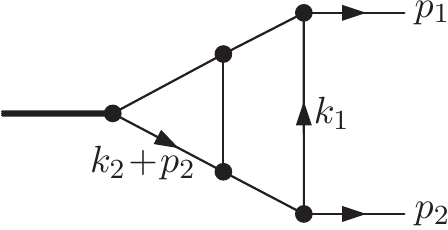}}\: 
\subfigure[]{\label{fig:2L_dlog_tri-b}\includegraphics[scale=0.67]{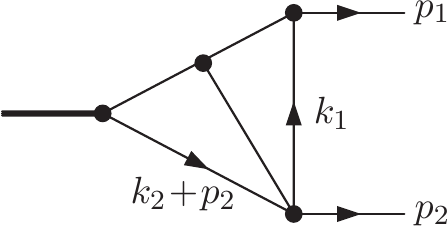}}\:
\subfigure[]{\label{fig:2L_dlog_tri-c}\includegraphics[scale=0.67]{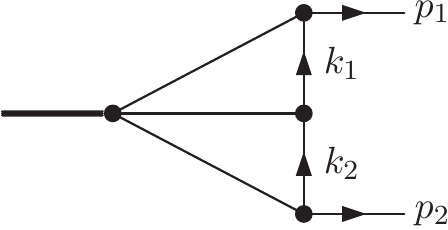}}\: 
\subfigure[]{\label{fig:2L_dlog_tri-d}\includegraphics[scale=0.67]{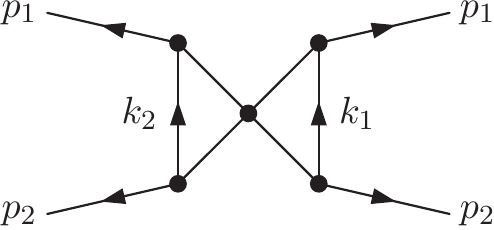}}
\end{center}
\caption{Representative two-loop planar triangles that admit a \dlog{} representation.}
\label{fig:2L_dlog_tri}
\end{figure}
In order to study all possible planar three-point Feynman integrands at two loops,
which correspond to Figs.~\ref{fig:parent_topos_Hgg-a}, \ref{fig:parent_topos_Hgg-b}, 
\ref{fig:parent_topos_Hgg-c}, 
it is convenient to use the following notation,
\begin{align}
{\cal I} =  \frac{d^4 k_1 \, d^4 k_2\, {\cal N} }{ k_{1}^{2} k_{2}^{2} \left(k_{1}-k_{2}\right)^{2} \left(k_{1}-p_{1}\right)^{2} \left(k_{2}-p_{1}\right)^{2} \left(k_{1}+p_{2}\right)^{2} \left(k_{2}+p_{2}\right)^{2} } \,,
\label{alltwoloopplanar}
\end{align}
to account for all possible propagators (see direction flow of the loop momenta in 
Fig.~\ref{fig:parent_topos_Hgg}). 

Different choices of ${\cal N}$ correspond to the integral families mentioned previously. The advantage of the formulation in terms of Eq.~\eqref{alltwoloopplanar} is that they can all be described simultaneously.

In order to determine all numerator factors ${\cal N}$ that lead to \dlog{} forms, 
we employ the  \Dlog{} program of \cite{Henn:2020lye}.
This readily gives us the following list of nine \dlog{} forms:
\begin{itemize}
\item two versions of Fig.~\ref{fig:2L_dlog_tri-a}, obtained by flipping $k_1\leftrightarrow k_2$, 
\item four versions of Fig.~\ref{fig:2L_dlog_tri-b}, obtained by flipping $k_1\leftrightarrow k_2$, 
and by flipping the graph along the vertical axis,
\item two versions of Fig.~\ref{fig:2L_dlog_tri-c}, obtained by flipping $k_1\leftrightarrow k_2$,
\item the squared triangle integrand of Fig.~\ref{fig:2L_dlog_tri-d}. 
\end{itemize}
Recall that we operate at integrand level, and hence, relations at integral level, e.g. sector symmetries and IBP identities, are not considered. 
Taking the latter into account would lead to four independent Feynman master integrals,
which correspond to the ones depicted in Fig.~\ref{fig:2L_dlog_tri}, 
but we work with the basis of nine \dlog{} integrands listed above.

\par\bigskip
Let us now turn to the computation of algebraic expressions for the 
\dlog{} integrands of Fig.~\ref{fig:2L_dlog_tri}. 
Thus, we begin the discussion with the ladder three-point scalar integrand
depicted in Fig.~\ref{fig:2L_dlog_tri-a}. 
We notice that, due to the way how this integrand is constructed, 
one can decompose it in terms of two one-loop integrands.
The first one corresponds to a scalar box 
with two off-shell external momenta,
whereas the second one, that is generated from the former, 
can be understood as a one-mass scalar triangle, 
\begin{align}
\omega_{\text{p},(a)}^{\left(2\right)} & = \mathcal{I}\left(\parbox{24mm}{\includegraphics[scale=0.5]{wtri2La.pdf}}\,s^2\right) \nonumber \\
& =
\mathcal{I}\left(\parbox{24mm}{\includegraphics[scale=0.5]{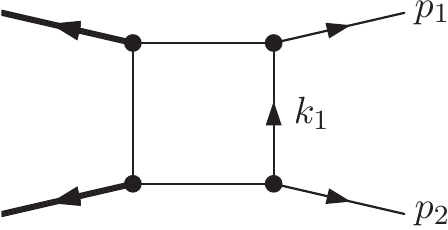}}\,k_2^2\,s\right)
\times
\mathcal{I}\left(\parbox{24mm}{\includegraphics[scale=0.5]{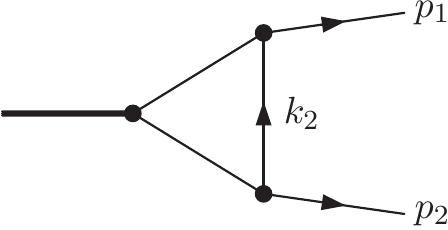}}\,s\right)
\,.
\label{eq:dlog_2L_ladder_tri}
\end{align}
Notice that in l.h.s. of Eq.~\eqref{eq:dlog_2L_ladder_tri}
we consider the integrand of Fig.~\ref{fig:2L_dlog_tri-a} with its appropriate
four-dimensional leading singularity, which was provided by \Dlog{}
and is needed to have a complete \dlog{} form for this integrand. 
Then, to follow a loop-by-loop approach, we start by considering
the subset of Feynman propagators, in the two-loop integrand, that depends on $k_1$.
These propagators generate a one-loop scalar box with two off-shell,
$-k_2-p_2$ and $k_2-p_1$,
and two on-shell, $p_1$ and $p_2$, momenta, whose \dlog{} form,
after taking into account the prefactor that comes from the leading singularity 
of the integrand, was considered in Eq.~\eqref{eq:twomasshardbox}. 
Notably, this prefactor, $s\,k_2^2$, allows to analyse the part of integrand 
that only contains $k_2$ but not $k_1$, in such a way that a one-mass
scalar triangle, together with its \dlog{} representation, 
is straightforwardly generated. 

Hence, the two-loop \dlog{} integrand $\omega_{\text{p},(a)}^{\left(2\right)}$
of Eq.~\eqref{eq:dlog_2L_ladder_tri} is expressed as a product of 
two one-loop \dlog{} integrands, 
\begin{align}
\omega_{\text{p},(a)}^{\left(2\right)}&=\omega^{\text{2mh}-\text{box}}\left(k_{1};p_{2},p_{1},k_{2}-p_{1}\right)\,\omega^{\text{1m}-\text{tri}}\left(k_{2};p_{1},p_{2}\right)\,,
\label{eq:triangle2a}
\end{align}
whose explicit algebraic expressions were given in Eqs.~\eqref{eq:1mass_tri_expr} 
and~\eqref{eq:twomasshardbox}.

A similar analysis and decomposition of the remaining 
two-loop \dlog{} integrands of Fig.~\ref{fig:2L_dlog_tri}
can be carried out with the help of the building blocks
discussed in the previous section.
We find,
\begin{align}
\omega_{\text{p},(b)}^{\left(2\right)} & =
\mathcal{I}\left(\parbox{24mm}{\includegraphics[scale=0.5]{wtri2Lb.pdf}}\,s\right) \nonumber\\
&= 
\mathcal{I}\left(\parbox{24mm}{\includegraphics[scale=0.5]{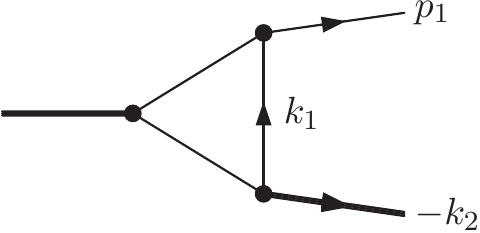}}\,(-2p_1\cdot k_2)\right)
\times
\mathcal{I}\left(\parbox{24mm}{\includegraphics[scale=0.5]{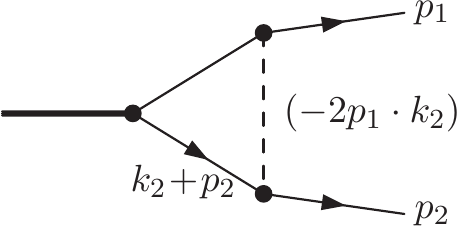}}\,s\right)\,, \nonumber\\
&=\omega^{\text{2m}-\text{tri}}\left(k_{1};p_{1},-k_{2}\right)\omega_{\text{Eikonal 1}}^{\text{1m}-\text{tri}}\left(k_{2};p_{1},p_{2}\right)\,,
\label{eq:triangle2b}
\\[1.5ex]
\omega_{\text{p},(c)}^{\left(2\right)} &=\mathcal{I}\left(\parbox{24mm}{\includegraphics[scale=0.5]{wtri2Lc.pdf}}\,s\right) \nonumber\\
&= 
\mathcal{I}\left(\parbox{24mm}{\includegraphics[scale=0.5]{wtri2Lb_tri1.pdf}}\,(-2p_1\cdot k_2)\right)
\times
\mathcal{I}\left(\parbox{24mm}{\includegraphics[scale=0.5]{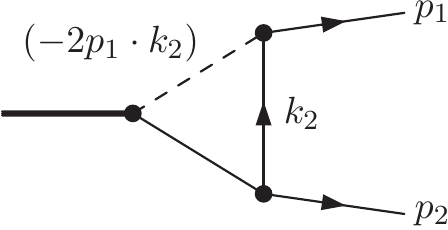}}\,s\right)\,, \nonumber\\
&=\omega^{\text{2m}-\text{tri}}\left(k_{1};p_{1},-k_{2}\right)\omega_{\text{Eikonal 2}}^{\text{1m}-\text{tri}}\left(k_{2};p_{1},p_{2}\right)\,,
\label{eq:triangle2c}
\\[1.5ex]
\omega_{\text{p},(d)}^{\left(2\right)} &=\mathcal{I}\left(\parbox{24mm}{\includegraphics[scale=0.5]{wtri2Ld.pdf}}\,s^2\right) \nonumber\\
&= 
\mathcal{I}\left(\parbox{24mm}{\includegraphics[scale=0.5]{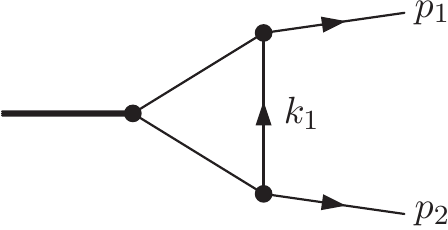}}\,s\right)
\times
\mathcal{I}\left(\parbox{24mm}{\includegraphics[scale=0.5]{wtri2La_tri.pdf}}\,s\right)\,,\nonumber\\
&=\omega^{\text{1m}-\text{tri}}\left(k_{1};p_{1},p_{2}\right)\omega^{\text{1m}-\text{tri}}\left(k_{2};p_{1},p_{2}\right)\,.
\label{eq:triangle2d}
\end{align}
The remaining \dlog{} forms for the full set of nine integrands are simply given by relabelling of loop momenta, as discussed above. 
Hence, for the purpose of illustrating the discussion in section~\ref{sec:applications}, 
we express $\omega_{\text{p},\left(i\right)}'$ as reflection of $\omega_{\text{p},\left(i\right)}$,
namely, $\omega_{\text{p},\left(i\right)}'=\omega_{\text{p},\left(i\right)}\big|_{k_{1}\leftrightarrow k_{2}}$. 

\begin{figure}[t]
\centering 
\subfigure[]{\label{fig:tri_np_dlog-a}\includegraphics[scale=0.67]{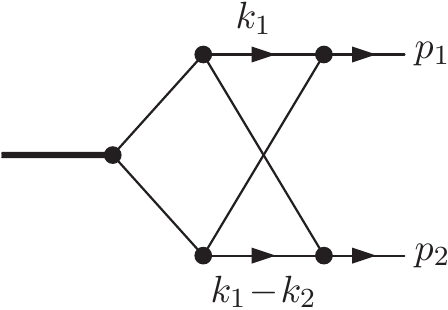}}\qquad 
\subfigure[]{\label{fig:tri_np_dlog-b}\includegraphics[scale=0.67]{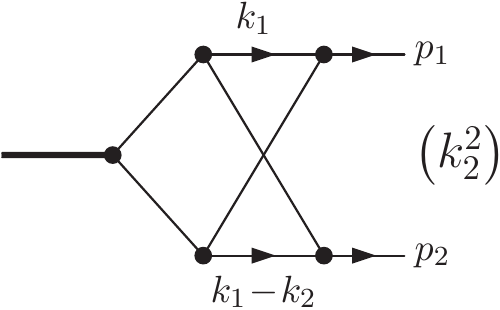}}
\caption{Two-loop non-planar triangles that admit a \dlog{} form.
The factor $(k_2^2)$ in the 
\dlog{} integrand~(b) corresponds
to an irreducible scalar product in the numerator. 
}
\label{fig:tri_np_dlog}
\end{figure}

Let us continue our analysis with \dlog{} integrands
that arise in non-planar sector of the three-point two-loop Feynman integrands.
This Feynman diagram has six propagators (see Fig.~\ref{fig:2L_dlog_tri-d}), 
and therefore  we look for \dlog{} integrands of the form,
\begin{align}
{\cal I} = \frac{d^4 k_1\,d^4 k_2\,{\cal N}} { k_{1}^{2} \left(k_{1}-p_{1}\right)^{2} \left(k_{2}-p_{1}\right)^{2} \left(k_{1}-k_{2}\right)^{2} \left(k_{1}-k_{2}-p_{2}\right)^{2} \left(k_{2}+p_{2}\right)^{2}} \,,
\label{eq:np_2Ltri}
\end{align}
for some (in general, loop-momentum-dependent) numerator ${\cal N}$.
We introduce the quantity $k_{2}^{2}$, in addition to the denominator factors 
in Eq.~\eqref{eq:np_2Ltri}, 
to express this numerator in terms of irreducible scalar products.
  
Thus, by making use of the built-in routines of \Dlog{}, we find eight \dlog{} integrands:
\begin{itemize}
\item two versions of Fig.~\ref{fig:2L_dlog_tri-b}, obtained by shrinking 
either the first or second propagator of~\eqref{eq:np_2Ltri},
\item two versions of the reflection Fig.~\ref{fig:2L_dlog_tri-b}, obtained by shrinking 
either the fourth or firth propagator of~\eqref{eq:np_2Ltri},
\item two versions of Fig.~\ref{fig:2L_dlog_tri-c}, obtained by shrinking 
either the third or sixth propagator of~\eqref{eq:np_2Ltri},
\item one version of the non-planar triangle of Fig.~\ref{fig:tri_np_dlog-a},
\item one version of Fig.~\ref{fig:tri_np_dlog-b}, which corresponds to the non-planar triangle 
with the irreducible scalar product $k_2^2$. 
\end{itemize}

Notice that in this \dlog{} basis, obtained from the non-planar sector 
through the integrand family~\eqref{eq:np_2Ltri}, 
six of these \dlog{} integrands were already considered in the planar sector.
This pattern was expected due to the isomorphisms between sub-topologies 
coming from planar and non-planar sectors, and 
leads to identities after integration. 
However, for our integrand analysis of this non-planar class of diagrams, 
it is important that we write down the planar integrals from the sub-sectors in the notation of Eq.~\eqref{eq:np_2Ltri} 
(for more details see Sec.~\ref{sec:decodlog}).

In the following, to simplify the notation for non-planar \dlog{} integrands
by keeping in mind the isomorphism between different sub-topologies, 
we introduce the shorthand notation
$\omega_{\text{np},(i);j}$ in the \dlog{} forms
to express that the loop topology of this integrand 
is isomorphic to the graph class $(i)$ in the planar sector (see Fig.~\ref{fig:2L_dlog_tri}) when
the $j$th propagator in the integrand family~\eqref{eq:np_2Ltri} is removed.

Since the \dlog{} expressions for the planar topologies can be obtained from the formulas given above by a simple change of variables,
let us focus on the genuine non-planar integrals of Fig.~\ref{fig:tri_np_dlog}.

We have the loop-by-loop decomposition,
\begin{align}\label{eq:triangleNPe}
\omega_{\text{np},(e)}^{\left(2\right)} &= 
\mathcal{I}\left(\parbox{24mm}{\includegraphics[scale=0.5]{wtri2L_np.pdf}}\,s^2\right)
\notag\\
 &=
\mathcal{I}\left(\parbox{24mm}{\includegraphics[scale=0.5]{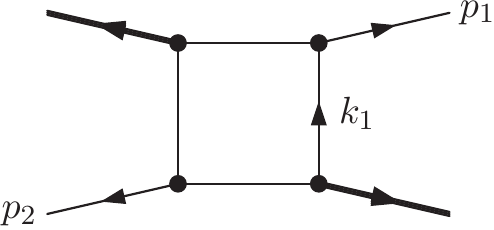}}\,
(2e_3\cdot k_2)(2e_4\cdot k_2)
\right)
\times
\mathcal{I}\left(\parbox{24mm}{\includegraphics[scale=0.5]{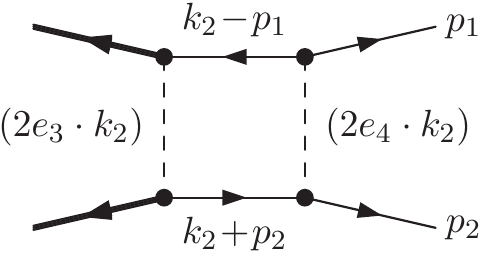}}\ s\right)
\notag\\
&=
\omega^{\text{2me}-\text{box}}\left(k;p_{1},-k_{2},p_{2}\right)
\omega_{\text{Eikonal 1}}^{\text{Box}}\left(k_{2};p_{1},p_{2}\right)\,,
\end{align}
and
\begin{align}\label{eq:triangleNPf}
\omega_{\text{np},(f)}^{\left(2\right)} &=
\mathcal{I}\left(\parbox{24mm}{\includegraphics[scale=0.5]{wtri2L_np.pdf}}\,s\,k_2^2\right)
\notag\\
&= \mathcal{I}\left(\parbox{24mm}{\includegraphics[scale=0.5]{wtri2L_np_box.pdf}}\,
(2e_3\cdot k_2)(2e_4\cdot k_2)
\right)
\times
\mathcal{I}\left(\parbox{24mm}{\includegraphics[scale=0.5]{wtri2L_np_tri.pdf}}\ s\,k_2^2\right) \nonumber\\
&=
\omega^{\text{2me}-\text{box}}\left(k;p_{1},-k_{2},p_{2}\right)
\omega_{\text{Eikonal 2}}^{\text{Box}}\left(k_{2};p_{1},p_{2}\right)\,.
\end{align}

To summarise, in this section, we studied the algebraic structure of \dlog{} integrands 
at one and two loops, focusing on integrals needed for the $gg\to H$ process, 
which includes planar and non-planar Feynman diagrams.
We found a complete list of \dlog{} integrands with the help of the algorithm \Dlog{}, and then provided, for each of the basis elements, expressions in a \dlog{} form. 
For studies at integrand level, a loop-by-loop approach turned out to be sufficient, and very useful. This approach allowed us to easily find \dlog{} forms with a single term only.

The main results of this section are the two-loop \dlog{} representations given in 
Eqs.~\eqref{eq:triangle2a}, \eqref{eq:triangle2b}, \eqref{eq:triangle2c}, \eqref{eq:triangle2d}, \eqref{eq:triangleNPe}, \eqref{eq:triangleNPf}. 
The complete list of expressions for the \dlog{} planar and non-planar integrand basis (9 and 8 integrands, respectively) is obtained from these by symmetry/reparametrisation, as explained above. These expressions are the main input for the maximal weight projection method that is discussed in section~\ref{sec:decodlog}.

\section{Maximal weight projection}
\label{sec:decodlog}

\subsection{Description of the method}

As discussed in the introduction, we are interested in the following general question:
can we extract the maximal transcendental weight piece of a quantity in perturbative quantum field theory, such as scattering amplitudes or correlation functions? 
Our starting point is an expression of the quantity of interest in terms of Feynman diagrams.
The idea is to analyse the expression for the Feynman integrand, i.e. prior to an (often computationally-heavy) analysis of relations between integrated functions. 

Recall that we expect the maximal weight of an $L$-loop integral in four dimensions to be $2L$~\cite{Hannesdottir:2021kpd}.
Moreover, the concept of weight can be applied to the Laurent series of integrals in dimensional regularisation, with $D=4-2\eps$. 
In this context, one assigns weight $-1$ to $\eps$~\cite{Henn:2013pwa}. In this way, a uniform and maximal weight function at $L$ loops is one where each term in the $\eps$ expansion has weight $2L$.

Based on experience from ${\cal N} =4$ sYM and related insights into differential equations that Feynman integrals satisfy, we expect that one can predict which terms in the integrand give maximal weight contributions, and which do not.
In the case of Feynman integrals evaluating to multiple poly-logarithms, which is our focus in this paper,
the idea is that Feynman integrals whose integrands can be written in \dlog{} form contribute to the maximal weight piece.
Therefore, the first step in our method is to obtain a basis of such \dlog{} integrands for the quantity of interest.
Let us denote the set of $m$ \dlog{} integrands at $L$ by 
\begin{align}
d\,{\rm log\; integrand\; basis: } \quad \{ {\cal I}^{(L)}_{i} \}\,,\quad i \in \{1,\ldots, m\} \,.
\label{eq:integrandbasis}
\end{align}
In addition to knowing what the basis elements are, we also need to know an explicit expression of their integrand in \dlog{} form, i.e. an expression of the form
\begin{align}
 {\cal I}^{(L)}_{i} = \sum_j b_{ij} \prod_{k=1}^{4L} d\log \alpha_{ijk} \,.
 \label{eq:integrandbasisform}
\end{align}
Here, we have assumed that we consider integrals in four dimensions, 
but one could also consider other integer dimensions.

To give an explicit example of~\eqref{eq:integrandbasisform}, consider the one-loop triangle integral from Eq.~\eqref{eq:1mass_tri_expr}. In this case, $L=1$, and the sum over $j$ in 
Eq.~\eqref{eq:integrandbasisform} involves a single term only. This fact is a special property that can sometimes be achieved (and is the case for all integrals considered in this paper). It will be useful in the following, but is not essential to our method.

One procedure for obtaining a basis~\eqref{eq:integrandbasis}, 
and the explicit representations~\eqref{eq:integrandbasisform} was discussed in section~\ref{sec:algebraic}, and a basis for planar and non-planar \dlog{} integrands relevant for the process $gg\to H$ at two loops was given. 
Remarkably, in the representations we found, the sum over $j$ in Eq.~\eqref{eq:integrandbasisform} has a single term only.

Let us assume that a basis and explicit expressions of \dlog{} integrands have been found.
Further, let $\cal{A}^{(L)}$ be the quantity we are interested in, for which we have a representation in terms of a loop integrand $\omega^{(L)}$,
\begin{align}
\mathcal{A}^{\left(L\right)} = \int \omega^{\left(L\right)}\,.
\end{align}
As mentioned in the introduction, we will consider $\omega$ in four dimensions, even if ${\cal A}$ is computed in $D=4-2\eps$ dimensions. Section~\ref{sec:evanescent} will address in more detail questions related to evanescent terms.\footnote{A related comment is that we do not expect the regularisation scheme to 
play a role when extracting the maximal transcendental weight
of a given scattering amplitude.}

The next step is then to imagine a decomposition of  $\omega^{\left(L\right)}$ in terms of our \dlog{} basis, plus other terms,
\begin{align}
\omega^{\left(L\right)} =\sum_{i=1}^{m}c_{i}\,\mathcal{I}_{i}^{\left(L\right)}+\hdots\,.
\label{eq:mysca}
\end{align}
The ellipses correspond to terms with at least a double pole. These are the terms that we wish
to drop when the projection onto the \dlog{} basis is performed.
 The reason is that we expect such terms to contribute less-than-maximal weight terms only.
In other words, given Eq.~(\ref{eq:mysca}) we define the projection operator ${\cal P}$ at integrand level as follows,
\begin{align}
{\cal P} \left( \omega^{\left(L\right)} \right) =\sum_{i=1}^{m}c_{i}\,\mathcal{I}_{i}^{\left(L\right)}\,.
\label{eq:projection}
\end{align}
Let us remark that in this decomposition
no relations at integral level are needed.
This is essential for the analysis we carry out throughout this paper. 

Taking into account Eq.~(\ref{eq:integrandbasisform}), the integrand 
of the amplitude $\omega^{\left(L\right)}$ must have the form
\begin{align}
{\cal P} \left(  \omega^{\left(L\right)}  \right)  & =\sum_{i=1}^{n} \tilde{c}_{i}\,\prod_{j=1}^{4L}\text{\dlog}\left(\tau_{i,j}\right)\,,
\label{eq:ampdeco}
\end{align}
with $\tau_{i,j}$ being some (often rational) functions depending on the loop components $\alpha$
of parametrisation~\eqref{eq:LoopParam}. 

Let us note that in r.h.s. of Eq.~\eqref{eq:ampdeco} in general, $n\geqslant m$. For instance, if we take $\omega$ to be the integrand of the  one-mass triangle, then we have $n=1$, but according to Eq.~\eqref{triangledlogalternative} we may have $m=4$. This is related to the fact that \dlog{} representations are by no means unique.

Our goal is to determine the $m$ coefficients $c_{i}$. 
As mentioned above, 
this is a problem that is very close to what is usually done when using generalised unitarity methods, see e.g.~\cite{Mastrolia:2006ki,Forde:2007mi}.
In fact, the set of $n$ coefficients $\tilde{c}_{i}$ contains the necessary information we are looking for. In practice, if we can compute a subset of $\tilde{c}_{i}$ whose map to the $c_i$ has rank $m$, then we can invert the system. 


In order to extract the coefficients $\tilde{c}_{i}$ in Eq.~(\ref{eq:ampdeco}), we make use of the Cauchy residue theorem, 
similar to the way how it is carried out in generalised unitarity. 
Thus, the locus where residues are extracted is determined once and for
all by the set of \dlog{} forms present in the decomposition of scattering amplitude.
Then, we can naively extract $\tilde{c}_{i}$ as, 
\begin{align}
\tilde{c}_{i} & =\oint_{\tau_{i,1}=0}\hdots\,\oint_{\tau_{i,4L}=0}\omega^{\left(L\right)}\,.
\label{eq:mycis}
\end{align}
Let us explain what we mean by `naively'.
The way residues are extracted in uni-variate
rational functions is unambiguous.
However, for multi-variate rational functions, the extraction of residues in a variable-by-variable approach, 
is more subtle. 
In particular, residues in $\tau_{i,j}=0$ do not commute.

This subtlety can, in principle, be overcome by employing the Grothendieck residue~\cite{Griffiths:433962},
however, for the calculations carried out in this paper, we make use of methods 
based on Gr\"obner basis and polynomial division.\footnote{For 
a pedagogical and detailed discussion of techniques to perform
multi-variate polynomial division, we refer the reader to Ref.~\cite{cox1992}.} 
Indeed, the computation of multi-variate residues~\cite{MR0463157,Griffiths:433962},
has been applied to the calculation of multi-loop scattering amplitudes in the context of generalised 
unitarity~\cite{Sogaard:2013fpa,Sogaard:2014ila,Sogaard:2014oka,Hauenstein:2014mda,Larsen:2017aqb,Bosma:2017ens}.
Let us also comment that the study of multi-variate polynomial division has been 
employed to perform reductions of multi-loop scattering amplitudes at 
integrand level~\cite{Mastrolia:2011pr,Zhang:2012ce,Badger:2013gxa,Mastrolia:2012an,Mastrolia:2012wf,Sogaard:2013fpa,Mastrolia:2016dhn,Mastrolia:2016czu}, inspired by the well-known Ossola-Papadopoulos-Pittau (OPP)
decomposition~\cite{Ossola:2006us}. 
In the present work, we rely on the 
partial-fraction decomposition introduced by 
Le\u{\i}nartas~\cite{leinartas1978,raichev2012},
which allows for an unambiguous extraction of certain 
coefficients of poles.
Improved versions of the Le\u{\i}nartas' decomposition have been recently provided in 
Refs.~\cite{Meyer:2016slj,Meyer:2017joq,Abreu:2019odu,Boehm:2020ijp,Heller:2021qkz},
where an extensive use of multivariate polynomial division,
or, said differently, reduction of polynomials,
through Gr\"obner basis is carried out.
In this paper, we use of the \Mathematica{} package \MultivariateApart{} provided together with Ref.~\cite{Heller:2021qkz}.

Before applying the method to a quantum field theory calculation, as an invitation let us first provide a toy example.

\subsection{Beta function toy example: an invitation}
\label{sec:toy1d}

In order to illustrate the decomposition of a given integrand in terms 
of \dlog{} forms as well as the extraction of the maximal weight contribution
of a given integrand,   
let us start with the one-dimensional example of the beta function. 
In view of its compact structure, one can easily cast interesting properties
and the main objective of this manuscript that in the following sections 
is extended to scattering amplitudes. 

Let us define the one-form
\begin{align}
\omega_{a,b}&=dz\,z^{-1-b+\eps}\left(1-z\right)^{-1-a+\eps}\,,
\label{defomegaab}
\end{align}
where $\eps$ is the analog of the dimensional regularisation parameter.
With this, we can define,
\begin{align}
I_{a,b}&=\int_{0}^{1}\omega_{a,b} \label{eq:beta}
=\frac{\Gamma\left(-a+\eps\right)\Gamma\left(-b+\eps\right)}{\Gamma\left(-a-b+2 \eps \right)}\,.
\end{align}
Thanks to the presence of $\eps$, the integral is well-defined for integer values of $a,b$.

In analogy with the dimensional regularisation case, let us now study \dlog{} forms for $\eps=0$.
We find that there are two forms, namely,
\begin{align}
\left\{ d\log z,-d\log\left(1-z\right)\right\}\,.
\label{eq:1d_dlogs}
\end{align}
These \dlog{} forms correspond to the integrands $\omega_{0,-1}$ and $\omega_{-1,0}$.

Due to a $z \leftrightarrow 1-z$ symmetry, we have that, after integration, $I_{a,b} = I_{b,a}$,
so that in particular $I_{-1,0} = I_{0,-1}$.
Inspecting the result~(\ref{eq:beta}), and expanding in $\eps$, 
we find,
\begin{align}
I_{0,-1}=I_{-1,0} 
& =  \frac{1}{\epsilon}-\frac{\pi^{2}}{6}\epsilon+2\,\zeta_{3}\,\epsilon^{2}+\mathcal{O}\left(\epsilon^{3}\right)
\,.
\label{eq:ut1d}
\end{align}
We see that all terms in the expansion in $\eps$ of $I_{-1,0}$ have uniform transcendental weight $1$, if one assigns weight $-1$ to $\eps$~\cite{Henn:2013pwa}.

Let us now consider a generic integrand that can be expressed in terms of $\omega_{a,b}$.
We would like to use our projection method to extract the leading weight contribution of the integral.
To do this, we apply decomposition~\eqref{eq:ampdeco} in terms of the 
\dlog{} integrands~\eqref{eq:1d_dlogs},
\begin{align}
\omega_{a,b}&=c_{0}\,\left[d\log z\right]+c_{1}\left[-d\log\left(1-z\right)\right]+\hdots\,,
\label{eq:deco_beta}
\end{align}
where once again ellipsis corresponds to terms with at least a double pole,
which are assumed to be irrelevant for the \dlog{} decomposition,
and $c_i$ are unknown coefficients that we need to determine.

It is clear from Eq.~(\ref{eq:deco_beta}) that $c_0$ is calculated by considering a contour integral around $z=0$. Applying this to the definition of $\omega_{a,b}$ in Eq.~(\ref{defomegaab}), we have,
 \begin{align}
c_{0}=\oint_{z=0}\omega_{a,b} |_{\eps=0} = \oint_{z=0}\frac{dz}{z^{a+1} \left(1-z\right)^{b+1}}=\binom{a+b}{a}
\,.
\end{align}
Likewise, $c_{1}$ is determined from a residue at $z=1$. In this particular
example, it turns to be $c_{1}=c_{0}$. 
Taking into account the relation $I_{-1,0} = I_{0,-1}$, we find the maximal weight projection 
\begin{align}
I_{a,b} &= 2\,  \binom{a+b}{a}  I_{-1,0} + { \rm weight \; drop \; terms } 
\notag\\
&= \binom{a+b}{a}\left(\frac{2}{\epsilon}-\frac{\pi^{2}}{3}\epsilon+4\,\zeta_{3}\,\epsilon^{2}\right) +\mathcal{O}\left(\epsilon^{3}\right) + { \rm weight \; drop \; terms } \,. 
\label{preditionweightdrop1dim}
\end{align}
In this specific case the leading weight piece, $I_{-1,0}$ has weight $1$, see Eq.~(\ref{eq:ut1d}).
This means that ``weight drop terms'' in this specific case are terms with transcendental weight $0$ or less.

In order to have a very explicit example, let us consider the values $m=5$ and $n=3$,
whose integrated expression, 
after expanding up-to second order in $\epsilon$, becomes, 
\begin{align}
I_{5,3}
=&
\frac{112}{\epsilon}-\frac{56}{3}\,\pi^{2}\,\epsilon+224\,\zeta_{3}\,\epsilon^{2}
\notag\\
&-\frac{2216}{15}-\frac{19468}{225}\epsilon+\epsilon^{2}\left(-\frac{234554}{3375}+\frac{1108}{45}\pi^{2}\right)+\mathcal{O}\left(\epsilon^{3}\right)\,.
\label{eq:beta_35}
\end{align}
Here, the first line of Eq.~\eqref{eq:beta_35} corresponds exactly to 
the maximal weight contribution, as predicted by our decomposition (i.e.
after setting $m=5$ and $n=3$ in Eq.~\eqref{preditionweightdrop1dim}),
while the terms in the second line all have less transcendental weight compared to the first line.

\section{Application: $H\to gg$ at two loops}
\label{sec:applications}

In the previous section, we provided a procedure to extract maximal leading 
transcendental weight terms of Feynman integrals.
This is done by applying a unique partial fractioning of the Feynman integrand via
Le\u{\i}nartas' decomposition method. 
In this way, given the knowledge of \dlog{} forms for a given multi-loop scattering amplitude,
an algebraic decomposition, as the one displayed in Eq.~\eqref{eq:ampdeco}, 
can straightforwardly be achieved. 

In this section we, wish to apply this new method to the 
two-loop scattering amplitude $H\to gg$ in the large top quark mass limit.  
This tests the method for the multi-variate decomposition in a non-trivial setting that involves both planar and non-planar Feynman diagrams.
We have already provided a basis of \dlog{} forms for this process in section~\ref{sec:decodlog}.
We now work out the \dlog{} decomposition of planar and non-planar integrands.

\begin{figure}[t]
\centering
\includegraphics[scale=0.7]{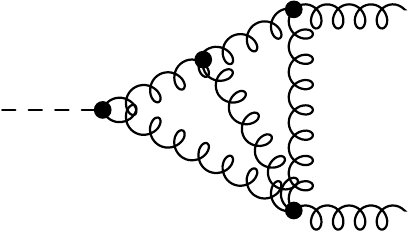}
\includegraphics[scale=0.7]{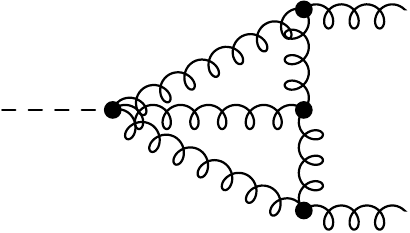}
\includegraphics[scale=0.7]{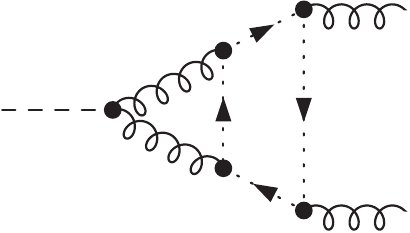}
\includegraphics[scale=0.7]{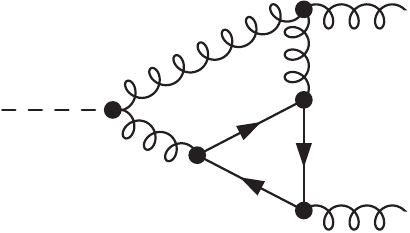}
\includegraphics[scale=0.7]{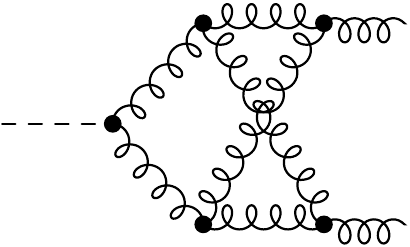}
\caption{Representative two-loop Feynman diagrams for the scattering amplitude $H\to gg$.
Curly, dashed, and straight lines represent, respectively, 
gluon, ghost, and massless fermions particles. 
 }
\label{fig:hgg}
\end{figure}

This scattering amplitude, because of Lorentz invariance, admits the decomposition, 
\begin{align}
\mathcal{A}_{Hgg}^{\left(2\right)}&=
g_\text{S}^4\,g_{\text{EFT}}\,\left(g^{\mu_{1}\mu_{2}}-\frac{2p_{1}^{\mu_{2}}p_{2}^{\mu_{2}}}{s_{12}}\right)
\delta^{a_1 a_2}
\varepsilon_{1,a_1}^{\mu_{1}}\varepsilon_{2,a_2}^{\mu_{2}}\,A_{1}^{\left(2\right)}\,,
\end{align}
with $g_{\text{S}}$ and $g_{\text{EFT}}$ stand for strong and 
$Hgg$-effective coupling constants, respectively, and $a_i$
correspond to colour indices. 

In order to carry out the calculation of $A_1^{(2)}$, 
we consider as internal particles: gluons, ghosts and massless fermions, 
as depicted for representative diagrams in Fig.~\ref{fig:hgg}.
The generation of Feynman diagrams and construction of their integrands were
carried out with aid of the \Mathematica{} packages \FeynArts~\cite{Hahn:2000kx}, 
\FeynCalc~\cite{Shtabovenko:2016whf,Shtabovenko:2016sxi},
and \FeynRules~\cite{Christensen:2008py}. 

Since our approach strongly relies on algebraic manipulations at integrand level, 
we perform a grouping of diagrams according 
to the propagators present in the integrand families~\eqref{alltwoloopplanar}
and~\eqref{eq:np_2Ltri}, for planar and non-planar sectors, respectively.
Within those, four parent topologies, or Feynman integrands with six propagators, 
are found and displayed in Fig.~\ref{fig:2L_dlog_tri}.

Thus, with the integrands obtained in the common set of propagators, 
we can project the latter onto the \dlog{} basis studied in section~\ref{sec:2Ldlog},
by following the procedure described in section~\ref{sec:decodlog}. 
In effect, for this form factor, we expect to have the decomposition, 
\begin{align}
\mathcal{P}\left(A_{1}^{\left(2\right)}\right)&=
\sum_{i=1}^{9}c_{\text{p},i}\,\omega_{\text{p},i}^{\left(2\right)}
+\sum_{i=1}^{8}c_{\text{np},i}\,\omega_{\text{np},i}^{\left(2\right)}\,.
\label{eq:deco_hgg}
\end{align}
Here $\mathcal{P}(\hdots)$ corresponds to the projection 
of a given integrand onto the two-loop \dlog{} basis,
and the labels $\text{`p'}$ and $\text{`np'}$ 
stand for planar and non-planar, respectively.
See section~\ref{sec:2Ldlog} for the definition of the nine planar and eight non-planar forms $\omega_{\text{p},i}$ and $\omega_{\text{np},i}$, respectively.

To proceed with the extraction of coefficients for each \dlog{} integrand,
we plug the parametrisation of loop momenta~\eqref{eq:LoopParam}
and expand out the various four-dimensional scalar products between internal and 
external momenta that appear in the form factor at integrand level. 
In other words, according to the notation of Eq.~\eqref{eq:ampdeco},
the integrand for this scattering amplitude becomes, 
\begin{align}
\omega^{(2)} = \mathcal{I}(D_s,s;\{\alpha\})\,.
\label{eq:int_tri2L}
\end{align}
Here, $D_s=(g_{[D_s]})^\mu_\mu$ corresponds to the dimension where 
internal particles, e.g. gluons, live in, and is related with the dimension
in the momentum integration as follows, 
\begin{align}
D_s = D+n_\epsilon = 4-2\epsilon+n_\epsilon\,.
\end{align}
In the four-dimensional helicity scheme, $D_s=4$ and $n_\epsilon=2\epsilon$,
while in conventional dimensional regularisation, $D_s=D$ 
and $n_\epsilon=0$~\cite{Gnendiger:2017pys}.

In Eq.~\eqref{eq:int_tri2L}, 
$\{\alpha\}=\{\alpha_{1,1},\hdots,\alpha_{1,4},\alpha_{2,1},\hdots,\alpha_{2,4},\}$
accounts for 
the set of loop momentum components, according to parametrisation~\eqref{eq:LoopParam}.
These algebraic manipulations
are performed with the aid of 
the \Mathematica{} packages \Sam~\cite{Maitre:2007jq} and \Tam~\cite{Bobadilla:2016mt}
that explicitly work out 
the various spinor products in terms of a reduced set of variables~\cite{Hodges:2009hk,Badger:2016uuq,TorresBobadilla:2017kpd}. 
For this particular example, we only need to deal with the single kinematic 
invariant $s=(p_1+p_2)^2$ and we set the dimension where internal particles live to four (cf. the discussion in section~\ref{sec:evanescent}).

We extract the coefficient of each \dlog{} integrand by appropriately taking 
the residue, 
making use of the package \MultivariateApart{}.

For the planar integrands, we find,
\begin{align}
\mathcal{P}\left(\parbox{24mm}{\includegraphics[scale=0.5]{hgg2L_8.pdf}}\right)
&
=\left[-2\omega_{\text{p},\left(a\right)}+\frac{3}{2}\left(\omega_{\text{p},\left(b\right)}+\omega_{\text{p},\left(b\right)}'\right)\right]s\,,
\notag\\
\mathcal{P}\left(\parbox{24mm}{\includegraphics[scale=0.5]{hgg2L_18.pdf}}\right)
&
=\left[-\frac{3}{4}\omega_{\text{p},\left(b\right)}-\frac{1}{2}\omega_{\text{p},\left(c\right)}\right]s\,,
\notag\\
\mathcal{P}\left(\parbox{24mm}{\includegraphics[scale=0.5]{hgg2L_13.pdf}}\right)
&
=\left[-\frac{3}{4}\omega_{\text{p},\left(b\right)}'-\frac{1}{2}\omega_{\text{p},\left(c\right)}\right]s\,,
\label{eq:proj_p_tri_2L}
\end{align}
whose algebraic expressions are reported in Sec.~\ref{sec:algebraic}.

The projections~\eqref{eq:proj_p_tri_2L} allow us to extract the maximal weight contributions 
in the planar sector, 
\begin{align}
\mathcal{P}\left(A_{1}^{\left(2\right)}\right)\Bigg|_{\text{planar}}&=\left[
-2\,\omega_{\text{p},\left(a\right)}+\frac{3}{4}\left(\omega_{\text{p},\left(b\right)}
+\omega_{\text{p},\left(b\right)}'\right)-\omega_{\text{p},\left(c\right)}
\right]s\,.
\label{eq:hgg_2L_planar}
\end{align}
Let us emphasis once again that in this decomposition no relations at integral 
level were involved and that is exactly the reason for the presence of the two \dlog{}
forms $\omega_{\text{p},\left(b\right)}$ and $\omega_{\text{p},\left(b\right)}'$, 
whose individual structure at integrand level is different, 
but after integration they turn out to be the same, 
\begin{align}
\int_{\ell_1,\ell_2}
\omega_{\text{p},\left(b\right)}'=
\int_{\ell_1,\ell_2}\omega_{\text{p},\left(b\right)}\,,
\label{eq:w1int_rel}
\end{align}
where, here and in the following, we use the shorthand notation,  
\begin{align}
\int_{\ell_{s}}\,\bullet\equiv e^{\left(4-D\right)\gamma_{E}/2}
\int\frac{d^{D}\ell_{s}}{\imath\,\pi^{D/2}}\,\bullet\,,
\end{align}
with $D=4-2\epsilon$.

Hence, with the additional relation~\eqref{eq:w1int_rel}, 
we can directly compare our result,
obtained purely in four dimensions, against the one that involves relations at 
integral level. To do so, we generate IBP identities
through the publicly available software \Reduze~\cite{vonManteuffel:2012np},
where, in the integrand constructed for the form factor, we substitute 
the integral relations generated by the latter and 
choose the master integrals that appear in Fig.~\ref{fig:2L_dlog_tri}. 
Thus, setting $D_s=4$ in the reduced integrand, 
we recover the very same result of Eq.~\eqref{eq:hgg_2L_planar}, 
after taking into account relation~\eqref{eq:w1int_rel}.

Let us now turn our attention to the non-planar sector,
in which the only non-vanishing Feynman diagram,
after projecting onto the \dlog{} basis for this sector
turns out to be,
\begin{align}
\mathcal{P}\left(A_{1}^{\left(2\right)}\right)\Bigg|_{\text{non-planar}}
&=
\mathcal{P}\left(\parbox{24mm}{\includegraphics[scale=0.5]{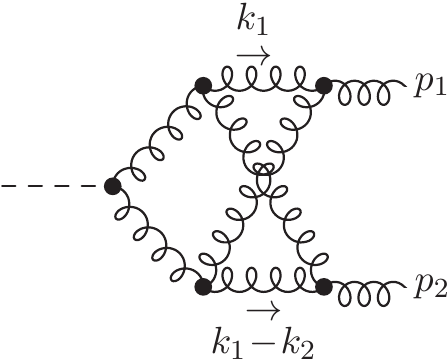}}\right)
\notag\\
&=\bigg[-\frac{1}{2}\omega_{\text{np},\left(e\right)}
-\frac{3}{8}\left(\omega_{\text{np},\left(b\right);1}
+\omega_{\text{np},\left(b\right);2}+\omega_{\text{np},\left(b\right);4}
+\omega_{\text{np},\left(b\right);5}\right)\notag\\
&\qquad+\frac{1}{2}\left(\omega_{\text{np},\left(c\right);3}
+\omega_{\text{np},\left(c\right);6}\right)\bigg]s\,.
\label{eq:hgg_2L_nonplanar}
\end{align}

Let us also notice that $\omega_{\text{np}(f)}$ 
does not appear in this \dlog{} decomposition
and it is due to cancellations, yet at integrand level, that arise from the 
generation of this amplitude. 
However, in view of this cancellation, it is worth elucidating in more details 
the decomposition of non-vanishing contributions of this \dlog{} form.
To do so, we consider a particular integrand where 
the presence of $\omega_{\text{np},(f)}$ is expected, 
\begin{align}
\mathcal{P}\left(\parbox{24mm}{\includegraphics[scale=0.5]{wtri2L_np.pdf}}\,s\,(k_2^2)^M\right)
&=\omega_{\text{np},\left(f\right)}\left(-s\right)^{M-1}\,,
\label{eq:tri2L_np_family}
\end{align}
with $M\in\mathbb{N}$. 

We emphasise that the decomposition~\eqref{eq:tri2L_np_family}
was carried out by following the procedure presented in Sec.~\ref{sec:decodlog}
and it was checked up-to $M=20$, which, for realistic computations 
of scattering amplitudes, one does not need to take into account such a  
high degree. 
Additionally, we made use of IBP identities to reduce the integrand 
(with up-to $M=4$)
in Eq.~\eqref{eq:tri2L_np_family}, to check that the four-dimensional projection
is in agreement with the former, after setting $D=4$.

Finally, we add up all contributions coming from planar and non-planar sectors, 
obtaining, 
\begin{align}
\mathcal{P}\left(A_{1}^{\left(2\right)}\right)=&\Bigg[-2\omega_{\text{p},\left(a\right)}-\frac{1}{2}\omega_{\text{np},\left(e\right)}
\notag\\
&-\frac{3}{8}\bigg(\omega_{\text{np},\left(b\right);1}+\omega_{\text{np},\left(b\right);2}+\omega_{\text{np},\left(b\right);4}+\omega_{\text{np},\left(b\right);5}-2\omega_{\text{p},\left(b\right)}-2\omega_{\text{p},\left(b\right)}'\bigg)
\notag\\
&+\frac{1}{2}\bigg(\omega_{\text{np},\left(c\right);3}+\omega_{\text{np},\left(c\right);6}
-2\omega_{\text{p},\left(c\right)}\bigg)\Bigg]\,s\,,
\end{align}
where, similar to the behaviour at integral level of the \dlog{} forms 
displayed in Eq.~\eqref{eq:w1int_rel}, we find an analogue for the non-planar \dlog{} integrands,
\begin{align}
\int_{\ell_1,\ell_2}
\omega_{\text{np},(i);j}=
\int_{\ell_1,\ell_2}
\omega_{\text{p},(i)}\,, 
\end{align}
that leads the form factor to be expressed as, 
\begin{align}
\int_{\ell_1,\ell_2}
\mathcal{P}\left(A_{1}^{\left(2\right)}\right)
& = 
\int_{\ell_1,\ell_2}
\Big[-2\omega_{\text{p},\left(a\right)}-\frac{1}{2}\omega_{\text{np},\left(e\right)}\Big]\,s\,,
\label{eq:uniform_2L_tri}
\end{align}
finding full agreement with approaches that rely on relations at integral level. 

Let us now put all results together. 
Since the main purpose of our procedure is to extract the maximal transcendental weight piece
of a given scattering amplitude, let us work this out explicitly.
Using the known expressions for the $\eps$-expansion of the master integrals (see e.g.~\cite{Gonsalves:1983nq,Kramer:1986sr}),
we find,
\begin{align}
\int_{\ell_1,\ell_2}
\mathcal{P}\left(A_{1}^{\left(2\right)}\right)
&=
\left(
-\frac{1}{\epsilon^{4}}+\frac{\pi^{2}}{12\epsilon^{2}}
+\frac{25\zeta_{3}}{6\epsilon}+\frac{7\pi^{4}}{120}\right)s
+\mathcal{O}\left(\epsilon\right)
\,.
\end{align}
This result is in perfect agreement
(up-to an irrelevant normalisation factor)
with the maximal leading terms of Eq. (17) of Ref.~\cite{Harlander:2000mg}.

Let us discuss this result. We see that the projection method gives us a preliminary insight of the structure of the 
scattering amplitude under consideration. In particular, thanks to the decomposition into \dlog{} forms, 
it is straightforward to extract the maximal transcendental weight contribution of a given scattering amplitude. 
The procedure proposed in Sec.~\ref{sec:decodlog} allows us
to unambiguously calculate multi-variate residues,
which is essential to extract the coefficients of \dlog{} integrands in the 
form factor decomposition~\eqref{eq:deco_hgg}. 
We find this procedure
very efficient.
In particular, because of the approach we follow is at integrand level,
the calculation of IBP identities in four dimensions 
is replaced by a multi-variate polynomial division modulo Gr\"obner basis.
Remarkably, contrary to the traditional generation of the former, 
there is no obstacle when increasing the rank in the numerator for a given 
integrand, as depicted in Eq.~\eqref{eq:tri2L_np_family} for a monomial with
degree 20 since the polynomial division is straightforwardly performed.

\section{Analysis of evanescent numerator terms}
\label{sec:evanescent}

In the previous sections, we discussed an algebraic procedure to extract
maximal transcendental weight of multi-loop scattering amplitude. 
In the present paper, we implemented this approach in a purely four-dimensional formulation,
as this has many practical advantages.
A natural and important question is to what extend subtleties can arise due to this approximation.
In this section, we address this in more detail.

In particular, one can write down evanescent integrand terms, by which we mean integrands that vanish 
in four dimensions, but are non-zero in $D=4-2\eps$ dimensions.
For example, given enough momenta in a given scattering process, such terms can be constructed 
in terms of certain Gram determinants. Closely related to this are `$\mu$'-terms, which depend on the 
difference between the $D$- and four-dimensional part of loop momenta.  
We wish to study in which situations such terms can impact the maximal weight projection.

We do so by analysing the one-loop case in detail, in subsection~\ref{sec:evanescent1}. 
Based on the conclusions drawn there, we then investigate four- and five-particle processes at two loops in subsection~\ref{sec:evanescent2}.

\subsection{One-loop analysis}
\label{sec:evanescent1}

To start this discussion, let us extend the four-dimensional parametrisation~\eqref{eq:LoopParam}
to $D$   dimensions, 
\begin{align}
k_{i\,\left[D\right]}^{\alpha} &= k_{i}^{\alpha}+k_{i\,\left[D-4\right]}^{\alpha}\,,
\label{eq:LoopParamD}
\end{align}
where we closely follow the notation of Ref.~\cite{Gnendiger:2017pys}. 
In this decomposition, 
$k_{i}^{\alpha}$  and $k_{i\,\left[D-4\right]}^{\alpha}$
live in independent sub-spaces, $k_{i}^{\alpha}\cdot k_{i\,\left[D-4\right]}^{\alpha}=0$,
and the extra dimensional product, or evanescent terms, 
are often defined as, $\mu_{ij}=-k_{i\,\left[D-4\right]}\cdot k_{j\,\left[D-4\right]}$.
In effect, the splitting $D$-dimensional loop momenta~\eqref{eq:LoopParamD}
has been extensively used in the four-dimensional helicity 
and 't Hooft-Veltman regularisation schemes, in which 
external momenta are always  cast in four dimensions. 

Analytic expressions for one-loop helicity amplitudes have been known for 
a long time. Interestingly, these amplitudes display a simple dependence
in terms of the so-called
evanescent terms $\mu_{11}$.
In effect, any $N$-point one-loop scattering amplitude 
in renormalisable  theories 
can be decomposed as follows~\cite{Giele:2008ve,Ellis:2008ir}, 
\begin{align}
\mathcal{A}_{N}^{\left(1\right)}
=&\sum_{i\in\text{pentagons}}e_{i}\,I_{5,i}^{\left(1\right),D}\left[1\right]
\notag\\
&+\sum_{i\in\text{boxes}}d_{i}\,I_{4,i}^{\left(1\right),D}\left[1\right]
+\sum_{i\in\text{triangles}}c_{i}\,I_{3,i}^{\left(1\right),D}\left[1\right]
+\sum_{i\in\text{bubbles}}b_{i}\,I_{2,i}^{\left(1\right),D}\left[1\right]
+\sum_{i\in\text{tadpoles}}a_{i}\,I_{1,i}^{\left(1\right),D}
\notag\\
&+\sum_{i\in\text{boxes}}d_{i,1}\,I_{4,i}^{\left(1\right),D}\left[\mu_{11}\right]
+\sum_{i\in\text{boxes}}d_{i,2}\,I_{4,i}^{\left(1\right),D}\left[\mu_{11}^{2}\right]
+\sum_{i\in\text{triangles}}c_{i,1}\,I_{3,i}^{\left(1\right),D}\left[\mu_{11}\right]
\notag\\
&+\sum_{i\in\text{bubbles}}b_{i,1}\,I_{2,i}^{\left(1\right),D}\left[\mu_{11}\right]\,,
\label{eq:1L_deco_mu11}
\end{align}
in which $I_{n,i}^{(1),D}$ corresponds to the 
$D$-dimensional 
$i$-th $n$-point Feynman integral
whose numerator, at integrand level, is given by the argument of the former.
For instance, a three-point topology with an evanescent terms in the numerator
is expressed as $I_{3,i}^{\left(1\right),D}[\mu_{11}]$.
The small letters, $\{a,b,\hdots,e\}$, are rational coefficients that 
only depend on the kinematics of the process under consideration
and not on the $\epsilon$ parameter.

As a consequence of working in $D=4-2\epsilon$ dimensions,  
one can relate scalar one-loop pentagon integrands as follows~\cite{Badger:2008cm,Melnikov:2010iu,Mastrolia:2010nb},
\begin{align}
\tilde{e}_i\,I_{5,i}^{(1),D}\left[1\right]
&=I_{5,i}^{(1),D}\left[\mu_{11}\right]+\sum_{i\in\text{boxes}}\tilde{d}_{i}\,I_{4,i}^{(1),D}\left[1\right]\,,
\label{eq:1L_5pt_a}
\end{align}
where, similar to coefficients $\{a,\hdots,e\}$ in decomposition~\eqref{eq:1L_deco_mu11}, 
$\tilde{d}_i$ and $\tilde{e}_i$ are rational  functions that only
depend on kinematics of the scattering process. 

Relation~\eqref{eq:1L_5pt_a} is straightforwardly obtained once the loop momentum 
components $\alpha$ are expressed in terms of the five denominators
as well as $\mu_{11}$. The latter is clearly manifest after performing a polynomial 
division modulo appropriate Gr\"obner basis.
Likewise, by following the latter argument, 
higher multiplicity scalar integrands, i.e. $N>5$, 
can always be reduced to pentagon integrands,
since denominators (or inverse of Feynman propagators)
are not independent in the former integrand. 

Let us now analyse the evanescent terms, i.e. those with $\mu_{11}$ numerators.
There are questions that we want to answer. First, can they produce leading weight terms,
i.e terms of weight two? 
Second, if so, at what order in $\eps$ do the leading weight terms appear?

To answer these questions recall that 
we can trade integrals with an insertion of $\mu_{11}$ for
higher-dimensional scalar integrals~\cite{Bern:1995db}. For one insertion of $\mu_{11}$, one has
\begin{align}
I_{n,i}^{\left(1\right),D}\left[\mu_{11}\right]=&- \eps \,I_{n,i}^{\left(1\right),D+2}\left[1\right]\,.
\label{eq:rel_mu11}
\end{align}
At this stage we can use knowledge of the transcendental weight of the six-dimensional integrals appearing on the r.h.s. of Eq.~\eqref{eq:rel_mu11}.
It turns out that the six-dimensional box integral has maximal weight two, while six-dimensional pentagon and hexagon integrals have maximal weight three. More generally, we expect six-dimensional integrals with $n\ge 5$ to have maximal weight three.
Taking into account the overall factor $\eps$ in Eq.~(\ref{eq:rel_mu11}), we conclude that $I_{n,i}^{\left(1\right),D}\left[\mu_{11}\right]$ has maximal weight one for $n=4$, and maximal weight two for $n\ge 5$. Moreover, since the six-dimensional integrals are finite, the evanescent terms contribute at $\cal{O}(\eps)$.

In other words, for $n=4$ the evanescent terms generated by $\mu_{11}$ in the numerator are irrelevant for the maximal weight projection.
For $n\ge 5$, this is no longer the case, unless one truncates the $\eps$ expansion at the finite part.

A similar analysis of transcendental weight drop in one-loop scalar integrals 
can be carried out by means of integrands constructed from Gram determinants. 
Let us define
\begin{align}
G\left(\begin{array}{c}
k_{1},\hdots,k_{s}\\
q_{1},\hdots,q_{s}
\end{array}\right)&\equiv\underset{i,j\in s\times s}{\text{det}}\left(2k_{i}\cdot q_{j}\right)\,,
\notag\\
G\left(k_{1},\hdots,k_{s}\right)&\equiv G\left(\begin{array}{c}
k_{1},\hdots,k_{s}\\
k_{1},\hdots,k_{s}
\end{array}\right)\,,
\end{align}
Then, a five-point integrand constructed from a Gram determinant,
after an analysis at integrand level~\cite{Melnikov:2010iu}, becomes,
\begin{align}
I_{5,i}^{(1),D}\left[G\left(k_{1},p_{1},p_{2},p_{3},p_{4}\right)\right]
&=-2\,G\left(p_{1},p_{2},p_{3},p_{4}\right)\,I_{5,i}^{(1),D}\left[\mu_{11}\right]\,.
\label{eq:gram1L}
\end{align}
We emphasise that at integrand level either side of Eq.~\eqref{eq:gram1L} vanishes 
in four dimension but yet gives contributions in $D$
with a weight drop in the transcendental degree, as previously discussed
for evanescent terms. 

In summary, we find that for one-loop scattering amplitudes with $n\ge 5$, the four-dimensional projection method may miss maximal weight terms starting from the ${\cal O}(\eps)$ terms. For example, at $n=5$, the pentagon integral~\eqref{eq:1L_5pt_a} with evanescent numerator $\mu_{11}$ starts as $\eps \times ({\rm weight\;three})$. 
This means that evanescent terms do not affect the maximal weight piece for $n$-point one-loop amplitudes up-to and including the finite part. The situation is even better for $n\le 4$ one-loop amplitudes, where evanescent terms do not lead to maximal weight pieces at all.

\subsection{Two-loop analysis at four and five points}
\label{sec:evanescent2}

In view of the results of the one-loop analysis of the previous section, it is interesting to think about higher loops.
In particular, we saw that evanescent terms in one-loop amplitudes can lead to maximal weight terms starting from ${\cal O}(\eps)$ for amplitudes with five or more external legs. In two-loop amplitudes, one may have a situation where evanescent terms in one loop are multiplied by divergences originating from another loop (most clearly this happens for products of one-loop amplitudes), and this can lead to maximal weight pieces that are missed, if the evanescent numerators are not kept. It is easy to write down examples of this. Does this mean that the four-dimensional projection method cannot be used starting from two loops and five points?

Of course, if this is the case, one could try to adapt the method by treating the $D$-dependence of integrands more carefully, as is done for example in Baikov representation~~\cite{Baikov:1996rk,Frellesvig:2017aai,Zeng:2017ipr,Bosma:2017hrk,Dlapa:2021qsl}.
However, this is clearly much more work, and it would be desirable to find a positive four-dimensional statement.
Hope comes from the fact that in most physical situations, at the end of the day, we wish to compute a (finite) result in four dimensions.
Moreover, both ultraviolet and infrared divergences are very well understood, in principle, and satisfy factorisation formulas.
Can we profit from these insights?

Let us discuss the divergences in turn. In the case of ultraviolet divergences, we heuristically know that they only involve less-than-maximal weight terms, so we can ignore them for the purpose of this discussion. The situation is different for infrared divergences, as they do involve maximal weight terms. So an infrared double pole, for example, when multiplied to the evanescent pentagon discussed in the last subsection, would produce a divergent maximal weight term at two loops. However, we know from general factorisation properties of infrared divergence that such terms are related to infrared counter-terms, multiplied by lower-loop amplitudes. For this reason, one can hope that such terms cancel when appropriate, infrared renormalised finite terms are considered. Such terms are sometimes called `hard functions' in the QCD literature~\cite{Collins:1989gx,Dixon:2008gr,Almelid:2015jia,Chicherin:2018old,TorresBobadilla:2020ekr}. 

In this subsection, we therefore discuss in some more detail the issue. We first study the four-point case, where, based on the above discussion, we do not expect any issue due to evanescent terms. Then we discuss an explicit example of a five-point amplitude in ${\cal N} =4$ super Yang-Mills that illustrates both the issue mentioned above, and its resolution.

\subsubsection*{Evanescent two-loop four-point terms have non-maximal weight}

We wish to analyse relevant evanescent terms for planar two-loop four-point amplitudes
(see Ref.~\cite{Gluza:2010ws} for closely related work).
Let us begin by writing down the most general integrand that could appear in a Yang-Mills theory.
In other words, we write all graphs with cubic vertices, 
and with at most one loop momentum per vertex. 
The numerator is then expressed as a linear combination of 
1316 monomials $(k_i\cdot q_j)$ with $j\in\{k_1,k_2,p_1,\hdots,p_3\}$.
Then, we project this integrand onto four dimensions 
through parametrisation~\eqref{eq:LoopParam}.
At this point, all scalar products between external and internal momenta 
are expressed in terms loop momentum components, 
namely, $(k_i\cdot q_i) \to \{\alpha\}$ with $q_i \in\{k_1,k_2,p_1,p_2,p_3\}$. 
By asking the latter projection to vanish, we find that our most generic vanishing 
four-dimensional integrand can be expressed in terms of 31 unknown coefficients. 
For example, one particular term in our ansatz is just the following Gram determinant, $G\left(k_{1},k_{2},p_{1},p_{2},p_{4}\right)$.
Let us call the evanescent numerator with the $31$ coefficients $N_{4}$.

\begin{figure}[t]
\begin{center}
\subfigure[]{\label{fig:2L_dlog_box-a}\includegraphics[scale=0.67]{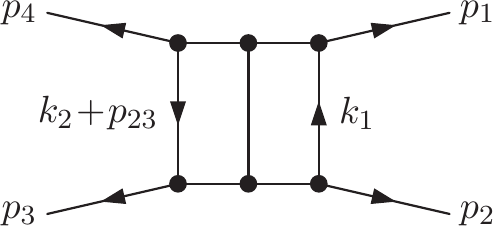}}\quad 
\subfigure[]{\label{fig:2L_dlog_box-b}\includegraphics[scale=0.67]{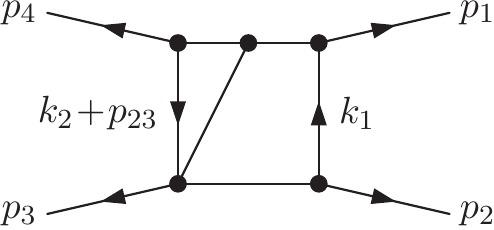}}\quad
\subfigure[]{\label{fig:2L_dlog_box-c}\includegraphics[scale=0.67]{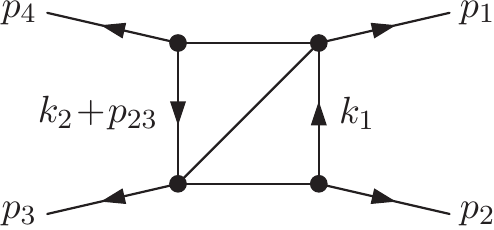}}\quad
\subfigure[]{\label{fig:2L_dlog_box-d}\includegraphics[scale=0.67]{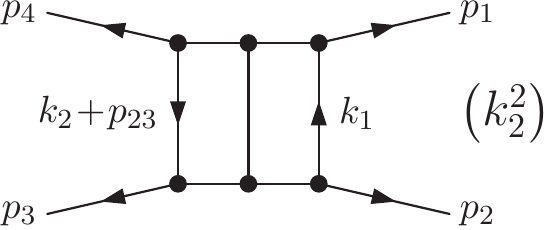}}
\end{center}
\caption{Representative two-loop planar boxes that admit a \dlog{} representation.}
\label{fig:2L_dlog_box}
\end{figure}

Next, we use IBP identities to reduce those integrands to a linear combination 
of eight master integrals. 
The latter are chosen such that they admit a \dlog{} representation~\cite{Wasser:2018qvj},
and correspond to the master integrals displayed in Figs.~\ref{fig:2L_dlog_tri} and~\ref{fig:2L_dlog_box}. 
Interestingly, we observe that the generic 
four-dimensional evanescent
satisfies
\begin{align}
\int_{k_{1},k_{2}} \frac{N_{4} }{\rm double\; box \;propagators}=&\epsilon\,\sum_{i=1}^{8}b_{i}(\eps) \,\omega_{i}^{\left(2\right)}\,,
\end{align}
with $b_{i}$ being rational functions in $\eps$ that are regular at $\eps=0$.
This equation allows us to conclude that the evanescent two-loop four-point terms lead to a weight drop. In other words, they are irrelevant for the maximal weight projection, at any order in $\eps$.

\subsubsection*{IR-renormalised five-point amplitude is free of ambiguities}

The planar ${\cal N}=4$ super Yang-Mills five-point amplitude~\cite{Cachazo:2006tj,Bern:2006vw} provides an instructive case of study of the effect of evanescent terms.

Let us write the amplitude as 
\begin{align}
A_{5} = A_{5; {\rm tree}}\,M_{5}\,,
\end{align}
where $A_{5; {\rm tree}}$ is the tree-level amplitude.
The loop factor, 
\begin{align}
M_{5} = 1 + g^2\,M_{5}^{(1)}  + g^4\,M_{5}^{(2)} + \ldots \,,
\end{align}
is a function that depends on kinematic invariants, 
on the coupling $g^2 = g_{\rm YM}^2 N_{c}/(16 \pi^2)$, and on the dimensional regulator $\eps$.

The one-loop amplitude $M_{5}^{(1)}$ consists of a sum of five infrared-divergence box integrals, plus exactly the evanescent pentagon integral~\eqref{eq:1L_5pt_a} discussed above. Therefore, this constitute an explicit field-theory example of the issue discussed above.
Our projection would give the correct result up-to and including the finite part.

In this particular case, there is additional structure that allows us to trace the effect of the evanescent terms. In fact, the one-loop evanescent terms are parity-odd, while all other one-loop terms are parity even. Interestingly, the duality between scattering amplitudes and Wilson loops states that~\cite{Alday:2007hr,Brandhuber:2007yx,Drummond:2007au,Drummond:2007aua}
\begin{align}\label{duality}
\log M_5 \sim \log W_5 + \cal{O}(\eps) \,,
\end{align}
where $W_{5}$ is a pentagonal Wilson loop. 
Here we do not wish to go into the details of this relation, but merely point out that the Wilson loop is parity even.
Therefore, if the duality relation~(\ref{duality}) holds (and this has been verified), then it implies something about the parity odd terms of the amplitude, and hence about the evanescent terms.
Expanding the l.h.s. of Eq.~\eqref{duality} in the coupling, we learn that apart from $M_{5}^{(1)}$, the following quantity must be parity even,
\begin{align}
M_{5}^{(2)} - \frac{1}{2} \left( M_{5}^{(1)}   \right)^2 + {\cal O}(\eps) \,.
\label{exponentiated2loop}
\end{align}
However, the product $\left( M_{5}^{(1)}   \right)^2$ clearly involves parity odd terms at $1/\eps$, as explained above.
Since the whole expression must be parity even, there must be corresponding parity odd terms in  $M_{5}^{(2)} $ to cancel the terms from the product. In other words, certain evanescent terms in $M_{5}^{(2)}$ cancel the unwanted evanescent terms coming from the product of lower-loop amplitudes, at least to ${\cal O}(\eps)$.\footnote{We remark that this is closely related to the observation that the same amplitude can be obtained using a purely four-dimensional construction of loop integrands, see for example Refs.~\cite{Arkani-Hamed:2010zjl} and~\cite{Eden:2010zz}. One possible interpretation of this is that the different integrands (with or without the evanescent terms, in this case) correspond to different scheme choices.} 
 
How can we interpret this result? The quantity $\log M$, or equivalently, the two-loop expression in Eq.~\eqref{exponentiated2loop} appears naturally when solving an infrared renormalisation group equation in a conformal field theory (see also Refs.~\cite{Anastasiou:2003kj,Bern:2005iz}). 
In other words, it takes care of the known partial exponentiation of non-Abelian soft and collinear divergences. In a more general, non-conformal field theory one could write down similar equations, that involve the $\beta$ function, that take case of infrared sub-divergences. 
Based on the observations made here, we find it likely that the four-dimensional projection gives the correct maximal weight terms, when applied to properly defined infrared-subtracted `hard' parts of scattering amplitudes.

\section{Conclusion and outlook}
\label{sec:conclusions}

In this paper, we proposed a method to extract the maximal weight contribution of scattering amplitudes.
The method does not require integration-by-parts identities, but operates directly at the level of the Feynman loop integrand.
It draws on insights that Feynman integrals with \dlog{} integrands evaluate to maximal weight functions.
For a given scattering process, the idea is to project the Feynman integrand onto a basis of such functions.
As such, the method combines well with generalised unitarity approaches.

After working out the method and discussing possible subtleties, we gave a proof-of-principle application, and extracted the maximal weight terms of the two-loop scattering amplitudes 
$gg\to H$ (in the heavy top-quark mass limit), 
which involves both planar and non-planar Feynman diagrams.
We found the method to work very efficiently.

For simplicity and convenience, we worked in this paper mostly with four-dimensional integrands, which have obvious advantages. We discussed the scope and limitations of the four-dimensional approach. In particular, we showed that certain evanescent terms are irrelevant for one-loop amplitudes up-to and including the finite part. We argued that the same may hold true when considering appropriately defined infrared subtracted finite parts of higher-loop amplitudes, and gave an explicit five-point two-loop example of this.

There are several interesting directions for further research:

1. A natural next step is to apply the method beyond the proof-of-principle cases presented here. 
Extracting the maximal weight terms of QCD amplitudes would allow to shed a first light on their analytic structure. 
For example, very recently, four-dimensional \dlog{} basis integrals have been presented for planar two-loop six-particle scattering processes~\cite{Bourjaily:2021hcp}, and it would be very interesting (keeping in mind however the discussion regarding evanescent terms) to obtain maximal weight expressions for six-gluon scattering in Yang-Mills theory, for example.

2. In some cases, one may be interested in extracting the maximal weight terms to all orders in the dimensional regulator $\eps$. In this case, one does need to keep evanescent terms. In order to do so, one needs to be able to perform the projections discussed in this paper, while keeping some information on the $(D-4)$-dimensional components of the loop momentum. A similar issue appears in the construction of uniform weight integrals in the context of the canonical differential equations method, and it was shown how to overcome it in Baikov representation~\cite{Baikov:1996rk,Frellesvig:2017aai,Zeng:2017ipr,Bosma:2017hrk,Dlapa:2021qsl}. 
It would be interesting to study our projection method in this setup.

3. The projection method requires the knowledge of a basis of \dlog{} integrands for a given scattering process. There is substantial knowledge and work on solving this problem~\cite{Lee:2014ioa,Meyer:2016slj,Prausa:2017ltv,Gituliar:2017vzm,Chicherin:2018old,Dlapa:2020cwj}, especially since this is also useful input for the canonical differential equations method~\cite{Henn:2013pwa}.
It may be interesting to generalise this in at least two directions. Firstly, as was observed in the literature, in many cases one can find \dlog{} representations that consist of a single term only. This is clearly desirable from the point of view of our projection method, but also beyond, see e.g. Ref.~\cite{Herrmann:2019upk} (direct integration). 
Can one understand when and how this property can be achieved. Secondly, we know that \dlog{} representations are closely related to geometric interpretations of scattering amplitudes, 
see e.g. Refs.~\cite{Arkani-Hamed:2013jha,Herrmann:2020qlt}. Is it possible to find a basis of \dlog{} integrands by geometric means? 

4. Finally, a further exciting direction is the extension of the projection method beyond leading weight. 
Amusingly, the most complicated terms, i.e. the leading weight terms, appear to be easy to obtain, while the less complicated terms appear to require more care to extract.
We have a good understanding of which integrands lead to maximal weight functions, and, conversely, which features of integrands lead to a weight drop. But can one quantify the weight drop property --- for example, if the leading weight terms have weight $2L$, is it possible to define a basis of next-to-maximal weight terms that have weight $2L-1$, and so on? This would provide a very useful new expansion.
One hint on how to identify such integrals could come from the canonical differential equations method~\cite{Henn:2013pwa}. In this approach it is especially clear that next-to-maximal weight integrals can be generated by taking a derivative of the leading weight terms (however, these integrals will have non-constant leading singularities).

\section*{Acknowledgments}
We thank Jaroslav Trnka for insightful discussions, 
and Pierpaolo Mastrolia and Yang Zhang for comments on the manuscript.
This research received funding from the European Research Council (ERC) under the European Union's Horizon 2020 research and innovation programme (grant agreement No 725110), {\it Novel structures in scattering amplitudes}.

\bibliographystyle{JHEP}
\bibliography{refs}
\end{document}